\documentclass[12pt]{article}
\usepackage{amsmath}
\usepackage{epsf}
\usepackage{epsfig}
\usepackage{here}
\usepackage{amssymb}
\usepackage{citesort}
\usepackage{graphicx}
\usepackage{latexsym}
\usepackage{axodraw}
\usepackage{bbm}

\newcommand{\be}{\begin{equation}}
\newcommand{\ee}{\end{equation}}
\newcommand{\bear}{\begin{eqnarray}}
\newcommand{\ear}{\end{eqnarray}}
\newcommand{\ba}{\begin{array}}
\newcommand{\ea}{\end{array}}
\def\la{\langle}
\def\ra{\rangle}
\def\der{\partial}

\makeatletter
\def\gsim{\compoundrel>\over\sim}

\def\compoundrel#1\over#2{\mathpalette\compoundreL{{#1}\over{#2}}}
\def\compoundreL#1#2{\compoundREL#1#2}
\def\compoundREL#1#2\over#3{\mathrel
         {\vcenter{\hbox{$\m@th\buildrel{#1#2}\over{#1#3}$}}}}
\makeatother

\def\l{\left}
\def\r{\right}

\begin{document}
  
\begin{center}

    {\Large\bf Non-minimal Split Supersymmetry}
    \\
    \vspace{0.8cm}
    \vspace{0.3cm}
    S.~V.~Demidov$^{a,}$\footnote{{\bf e-mail}: demidov@ms2.inr.ac.ru}, 
    D.~S.~Gorbunov$^{a,}$\footnote{{\bf e-mail}: gorby@ms2.inr.ac.ru}
    \\
    
    $^a${\small{\em 
        Institute for Nuclear Research of the Russian Academy of Sciences, }}\\
      {\small{\em
          60th October Anniversary prospect 7a, Moscow 117312, Russia
      }
      }
      \\
  \end{center}
  \vspace{2cm}
  \begin{abstract}
We present an extension of the minimal split supersymmetry model,
which is capable of explaining the baryon asymmetry of the Universe. 
Instead of MSSM we start from NMSSM and split its spectrum in such a
way that the low energy theory contains neutral particles, in addition 
to the content of minimal split supersymmetry. They trigger the strongly 
first order electroweak phase transition (EWPT) and provide an
additional source of CP-violation. In this model, we estimate the
amount of the baryon asymmetry produced during EWPT, using WKB 
approximation for CP-violating sources in diffusion equations. We also 
examine the contribution of CP-violating interactions to the electron
and neutron electric dipole moments and estimate the production of the 
neutralino dark matter. We find that both phenomenological and
cosmological requirements can be fulfilled in this model.
    
  \end{abstract}

\vskip 1cm


\newpage
\section{Introduction}
\label{intro_section}
In spite of approximate symmetry between particles and antiparticles,
our visible Universe is asymmetric in baryons. An explanation of this
fact ought to be addressed in any viable theory. Quantitatively, the
measurements of the anisotropy of the cosmic microwave background
constrain the baryon-to-photon ratio \cite{Bennett:2003bz} to be in
the range
\be
  \label{BAU}
  6.1\times 10^{-10} < \frac{n_{B}}{n_{\gamma}} < 6.9\times 10^{-10}.
\ee
Three necessary conditions (Sakharov conditions) must be fulfilled in
the early Universe to produce the baryon asymmetry
\cite{Sakharov:1967dj}: baryon number violation, C- and CP-violation
and departure from thermal equilibrium.
  
Electroweak baryogenesis \cite{Cohen:1993nk} is one of the most
interesting scenarios of baryon asymmetry generation. If the
electroweak phase 
transition (EWPT) is of the first order, it provides departure from
thermal equilibrium in cosmic plasma, as it proceeds through
nucleation of bubbles of the broken phase, their subsequent growth and
percolation.  The baryon asymmetry generated during the phase transition can be
washed out by rapid sphaleron processes in the broken phase
\cite{Kuzmin:1985mm}. In the Standard Model, the condition    
\be
\label{critical}
  \frac{v_{c}}{T_{c}}\gsim 1.1
\ee
(where $T_{c}$ is the critical temperature and $v_{c}$ is the Higgs
expectation value at this temperature) would ensure that the sphaleron 
transitions are suppressed inside the bubbles of the new phase
\cite{Moore:1998sw} and the baryon asymmetry survives. Nonperturbative  
analysis \cite{Kajantie:1996mn} revealed that, with the present 
experimental bound on the Higgs boson mass, the electroweak phase
transition in the Standard Model not only does not yield
(\ref{critical}), but is definitely not of the first order. Moreover,
in spite of the presence of CP-violating phase in CKM  
matrix, its contribution to the production of the baryon asymmetry is
too small to explain the observed value (\ref{BAU}). The Standard
Model fails to support electroweak baryogenesis. 

Supersymmetry is one of the most attractive ways beyond the framework
of the Standard Model. One recalls the cancellation of quadratic
divergences and gauge coupling unification among the main reasons for
interest in supersymmetric theories. Minimal Supersymmetric Standard
Model (MSSM) and its extensions suggest the lightest supersymmetric
particle (LSP) as a natural dark matter candidate and provide
mechanisms to generate the baryon asymmetry (for recent studies see 
Refs.~\cite{Carena:2002ss}, \cite{Huber:2000mg},
\cite{Konstandin:2005cd}, \cite{Huber:2006wf}
and references therein).  
  
Recently, motivated by landscape of vacua in string theory and
cosmological constant problem, models with split supersymmetry
have been proposed \cite{Arkani-Hamed:2004fb,Giudice:2004tc}. 
The spectrum of these models is governed by two scales. While the
electroweak scale $m_{ew}$ determines the masses of the Standard Model
particles and additional fermionic degrees of freedom (higgsinos and
gauginos) the other MSSM particles (scalars) have masses of order 
of a splitting scale $m_{s}$, which generally may be in a wide range
of $10^{4} - 10^{15}$~GeV. Such a pattern ensures the gauge coupling
unification, so that split SUSY can be incorporated into a GUT.  

Split SUSY enables one to avoid phenomenological difficulties with
CP- and flavor violation and also provides natural dark matter
candidates. But from the point of view of the electroweak 
baryogenesis\footnote{
For a discussion of prospects of the Affleck-Dine mechanism of
baryogenesis in split SUSY see Ref.~\cite{Kasuya:2005sb}.
},
split supersymmetry, in its minimal version, shares some disadvantages
of SM: the electroweak phase transition remains too weak, because the
additional, weakly coupled fermions 
do not strengthen EWPT. It is of interest to find a viable scenario,
which would be compatible with general split supersymmetry framework
and would provide conditions required for effective electroweak 
baryogenesis. One way, proposed in Ref.~\cite{Carena:2004ha}, is to
spoil the supersymmetric relations between the gauge and Yukawa
gaugino-higgsino couplings. 
For rather large Yukawas, the charginos and neutralinos are
light in the symmetric phase but heavy in the broken phase, so they
decouple from the cosmic plasma at the moment of the phase transition. 
During  the EWPT the particle species to be decoupled transfer their
entropy to the cosmic plasma, that results in the increase of the
temperature and the delay of the transition, providing a possibility to
satisfy the inequality (\ref{critical}).   


In this paper we pursue another approach and propose a model which
keeps all phenomenologically interesting properties of the minimal
split supersymmetry and at the same time is capable of producing the
baryon asymmetry at EWPT.  Instead of using MSSM as a starting point,
we begin with Non-Minimal  Supersymmetric Standard Model (NMSSM) and
split its spectrum in such a  way as to provide the low energy theory
with all features required for successful electroweak
baryogenesis. Indeed, additional neutral scalar fields, which are left
at low energies after the spectrum splitting, change zero temperature
potential and can make EWPT stronger. The model contains new sources
of CP-violation as 
well. Apart from their crucial role in the generation of the required
amount of the baryon asymmetry (\ref{BAU}), they also contribute to
the electric dipole moments (EDMs) of electron and neutron at the
level detectable in the near future experiments. Also, we study the
possibility that the lightest neutralino is dark matter particle. We
find that along with gaugino- and higgsino-like candidates there is a
region of parameter space where dark matter is mostly singlino. 

The outline of this paper is as follows. Sec.~\ref{model_section} 
contains the description of our model and its main 
features. In Sec.~\ref{ewpt_section} we study the strength of the 
electroweak phase transition and show that EWPT can be the strongly
first order. In Sec.~\ref{bau_section} we describe the baryon
asymmetry calculation procedure and present numerical results. In
Sec.~\ref{edm_section} we estimate the values of electron and neutron
EDMs in our model. Neutralino dark matter phenomenology is discussed
in Sec.~\ref{dark_matt} and Sec.~\ref{concl_section} contains our
conclusions.

\section{The model}
\label{model_section}
  
We begin with the most general NMSSM. The relevant for our study part 
of the superpotential is~\cite{Davies:1996qn}
\be
\label{superpotential}
  W = \lambda \hat{N}\hat{H}_{u}\epsilon\hat{H}_{d} + 
  \frac{1}{3}k\hat{N}^{3} + 
  \mu\hat{H}_{u}\epsilon\hat{H}_{d} +
  r\hat{N}.
\ee
Here\footnote{
  We denote superfields and their scalar components by letters with
  hat and without hat, respectively.
}
$\hat{H}_{u}$ and $\hat{H}_{d}$ are the Higgs doublets, $\hat{N}$ is a  
chiral superfield, which is a singlet with respect to the SM gauge
group and $\epsilon$ is antisymmetric $2\times 2$ matrix,
$\epsilon_{12} = 1$. Soft SUSY breaking terms for this model read
\begin{gather}
\label{bpotential}
  V_{soft} = \l(\lambda A_{\lambda}NH_{u}\epsilon H_{d} + 
  \frac{1}{3}kA_{k}N^{3} + 
  \mu BH_{u}\epsilon H_{d} + A_{r}N + h.c.\r) \\
+ m_{u}^{2}H_{u}^{\dagger}H_{u} +
m_{d}^{2}H_{d}^{\dagger}H_{d} + m_{N}^{2}|N|^{2}. \nonumber
\end{gather}
Electroweak baryogenesis in NMSSM has been explored in
Refs.~\cite{Huber:2000mg} in the context of low energy supersymmetry
breaking. We consider this model in the framework of split SUSY. For
this purpose some parameters need to be fine-tuned, like in the
minimal split SUSY, in such a way that the particle spectrum is split
into two parts. To strengthen the electroweak phase transition, some 
scalars can be recruited \cite{Anderson:1991zb}. To preserve gauge
coupling unification, these particles should not contribute to the
beta functions (or their contributions should be canceled) at least
at the leading order. So in the minimal case, the low energy theory
can contain (besides the usual split SUSY particles) singlets with
respect to the SM gauge group.

We consider the theory in which the only new source of explicit
CP-violation is $\mu$-parameter which for concreteness we take to be
exactly imaginary. All other parameters (including $B\mu$-term) of the
theory  
(\ref{superpotential}), (\ref{bpotential}) are taken to be real
(except in the analysis of the electron and neutron EDMs). This is
a purely technical assumption which we make in order that the parameter
space to be tractable. 

We examine splitting in the $(H_{u},H_{d},N)$ sector only, because
decoupling of 
squarks and sleptons is supposed to be similar to the case of the
minimal split SUSY. The tree level scalar potential of NMSSM is
\be
\label{potential}
V = V_{D} + V_{F} + V_{soft}\;,
\ee
where
\begin{gather}
  V_{D} = \frac{g^{2}}{8}\l(H^{\dagger}_{d}\sigma_{a}H_{d} +
  H^{\dagger}_{u}\sigma_{a}H_{u}\r)^{2} + 
  \frac{g^{\prime 2}}{8}\l(|H_{d}|^{2} - |H_{u}|^{2}\r)^{2}, \\
  V_{F} = 
  |\lambda H_{u}\epsilon H_{d} + kN^{2} + r|^{2} + 
  |\lambda N + \mu|^{2}\l(H_{u}^{\dagger}H_{u} 
  + H_{d}^{\dagger}H_{d}\r), 
\end{gather}
and the Higgs doublets are $H_{u} = \l(H_{u}^{+}, H_{u}^{0}\r)^{T}$
and $H_{d} = \l(H_{d}^{0}, H_{d}^{-}\r)^{T}$. Let us choose the vacuum
expectation values for neutral scalars as follows,
\[
\langle H_{d}^{0}\rangle = \frac{1}{\sqrt{2}}v_{d},\;\;\;
\langle H_{u}^{0}\rangle = \frac{1}{\sqrt{2}}v_{u}e^{i\phi_{H}},\;\;\;
\langle N\rangle = \frac{1}{\sqrt{2}}\l(v_{S} + iv_{P}\r) 
\equiv \frac{1}{\sqrt{2}}\sqrt{v_{S}^{2} + v_{P}^{2}}e^{i\phi_{S}}.
\]
Here we have taken into account that the effective potential depends
only on the sum of the phases of vev's of the Higgs doublets and one
of these phases can be set equal to zero by a gauge
transformation. The stationarity condition with respect to the phase
$\phi_{H}$, $\partial V/\partial\phi_{H} = 0$,
gives the following relation
\be
\label{phiH}
\sin{\phi_{H}} = -
\frac{\frac{k\lambda}{2}\l(v_{S}^{2} +
  v_{P}^{2}\r)\sin{(\phi_{H} - 2\phi_{S})} + 
\frac{\lambda A_{\lambda}}{\sqrt{2}}
\sqrt{v_{S}^{2} + v_{P}^{2}}
\sin{(\phi_{S} + \phi_{H})}}
{B\mu + \lambda r}.
\ee
To split the spectrum of the theory, we take soft SUSY breaking
parameters $B\mu$, $m_{u}^{2}$ and $m_{d}^{2}$ of the order of the
splitting scale $m_{s}^{2}$ and all other parameters in the scalar
sector (including all vev's) of the order of the electroweak scale
$m_{ew}$. Equation~(\ref{phiH}) implies the estimate $|\phi_{H}| \sim 
\l(m_{ew}/m_{s}\r)^{2}$ and therefore this phase can be safely
neglected in the following analysis. 

We find that vev's of the scalar fields at the minimum of the
potential (\ref{potential}) must satisfy the following conditions  
\begin{eqnarray}
m_{u}^{2} & = &
\frac{\bar{g}^{2}}{8}v^{2}\cos{2\beta}
- \frac{\lambda^{2}}{2}v^{2}\cos^{2}{\beta}
-\frac{\lambda^{2}}{2}v_{S}^{2} 
-\l(|\mu| + \frac{\lambda}{\sqrt{2}}v_{P}\r)^{2}
+ m_{A}^{2}\cos^{2}{\beta}, \label{1.1} \\
m_{d}^{2} & = &
- \frac{\bar{g}^{2}}{8}v^{2}\cos{2\beta}
- \frac{\lambda^{2}}{2}v^{2}\sin^{2}{\beta}
- \frac{\lambda^{2}}{2}v_{S}^{2}
- \l(|\mu| + \frac{\lambda}{\sqrt{2}}v_{P}\r)^{2} 
+ m_{A}^{2}\sin^{2}{\beta},  \label{1.2}  \\
m_{N}^{2} + 2kr & = &
\frac{1}{2}k\lambda v^{2}\sin{2\beta} 
- k^{2}\l(v_{S}^{2} + v_{P}^{2}\r)
-\frac{\lambda^{2}}{2}v^{2}
+\frac{\lambda
  A_{\lambda}}{2\sqrt{2}}\cdot\frac{v^{2}}{v_{S}}\sin{2\beta}
\label{1.3} \\
& & -\frac{kA_{k}}{\sqrt{2}}\l(v_{S} - \frac{v_{P}^{2}}{v_{S}}\r)
-\sqrt{2}\frac{A_{r}}{v_{S}}, \nonumber\\
m_{N}^{2} - 2kr & = &
-\frac{1}{2}k\lambda v^{2}\sin{2\beta}
-k^{2}\l(v_{S}^{2} + v_{P}^{2}\r)
-\frac{\lambda^{2}}{2}v^{2} 
- \frac{\lambda |\mu|v^{2}}{\sqrt{2}v_{P}}
+ \sqrt{2}kA_{k}v_{S}v_{P}, 
\label{1.4}
\end{eqnarray}
where we use the notations $\tan{\beta} =v_{u}/v_{d}$, $v^{2} =
v_{u}^{2} + v_{d}^{2}$, $\bar{g}^{2} = g^{2} + g^{\prime 2}$ and 
\[
m_{A}^{2} = \frac{2}{\sin{2\beta}}\l(\lambda r +
\frac{k\lambda}{2}\l(v_{S}^{2} - v_{P}^{2}\r) +
\frac{1}{\sqrt{2}}\lambda A_{\lambda}v_{S} + \mu B\r). 
\]
We expand the scalar fields about their vev's as follows,
\begin{gather}
{\rm Re} H_{d}^{0} = \frac{1}{\sqrt{2}}\l(v_{d} + h^{0}\cos{\beta} - 
H^{0}\sin{\beta}\r),\;\;
{\rm Im} H_{d}^{0} = \frac{1}{\sqrt{2}}\l(A^{0}\sin{\beta} +
G^{0}\cos{\beta}\r), \nonumber\\
{\rm Re} H_{u}^{0} = \frac{1}{\sqrt{2}}\l(v_{u} + h^{0}\sin{\beta} +
H^{0}\cos{\beta}\r),\;\;
{\rm Im} H_{u}^{0} = \frac{1}{\sqrt{2}}\l(A^{0}\cos{\beta} -
G^{0}\sin{\beta}\r),\nonumber\\
H_{d}^{-} = G^{-}\cos{\beta} + H^{-}\sin{\beta},\;\;
H_{u}^{+} = H^{+}\cos{\beta} - G^{+}\sin{\beta},\nonumber\\
N = \frac{1}{\sqrt{2}}\l(v_{S} + S + i(v_{P} + P)\r).\nonumber
\end{gather}
Here $G^{\pm}$ and $G^{0}$ are Goldstone modes which are eaten, due to 
the Higgs mechanism, by $W^{\pm}$- and $Z^{0}$-bosons, respectively,
$h^{0}$ and $H^{0}$ are the neutral scalar Higgs bosons, $A^{0}$ is
the Higgs pseudoscalar, $S$ and $P$ are singlet scalar and
pseudoscalar fields. 

Let us choose the basis in the space of neutral scalars as
$(h^{0},H^{0},A^{0},S,P)$. By making use of the conditions
(\ref{1.1}) - (\ref{1.4}), the elements of $5\times 5$ squared mass
matrix for neutral scalars read as follows, 
\bear
M_{11}^{2} & = & \frac{\bar{g}^{2}v^{2}}{4}\cos^{2}{2\beta} + 
\frac{\lambda^{2}v^{2}}{2}\sin^{2}{2\beta}, \\
M_{22}^{2} & = & m_{A}^{2}- \frac{\lambda^{2}v^{2}}{2}\sin^{2}{2\beta} + 
\frac{\bar{g}^{2}v^{2}}{4}\sin^{2}{2\beta}, \\
M_{33}^{2} & = & m_{A}^{2} ,\\
M_{44}^{2} & = & 2k^{2}v_{S}^{2} +
\frac{\lambda A_{\lambda}}{2\sqrt{2}}\frac{v^{2}}{v_{S}}\sin{2\beta}
+\frac{kA_{k}}{\sqrt{2}}\l(v_{S} + \frac{v_{P}^{2}}{v_{S}}\r)
-\sqrt{2}\frac{A_{r}}{v_{S}},
\ear
\bear
M_{55}^{2} & = & 2k^{2}v_{P}^{2} -
\frac{\lambda}{\sqrt{2}}\frac{|\mu|v^{2}}{v_{P}},\\
M_{12}^{2} & = & M_{21}^{2} = \l(\frac{\lambda^{2}}{4} - 
\frac{\bar{g}^{2}}{8}\r)v^{2}\sin{4\beta},\\
M_{13}^{2} & = & M_{31}^{2} = -k\lambda v_{S}v_{P} +
\frac{\lambda A_{\lambda}}{\sqrt{2}}v_{P}, \\
M_{14}^{2} & = & M_{41}^{2} = -k\lambda v_{S}v\sin{2\beta} +
\lambda^{2}v_{S}v -\frac{\lambda A_{\lambda}}{\sqrt{2}}v\sin{2\beta}, \\ 
M_{15}^{2} & = & M_{51}^{2} = k\lambda v_{P}v\sin{2\beta} +
\sqrt{2}\lambda v\l(|\mu| + \frac{\lambda}{\sqrt{2}}v_{P}\r), \\
M_{23}^{2} & = & M_{32}^{2} = 0, \\
M_{24}^{2} & = & M_{42}^{2} = -k\lambda v_{S}v\cos{2\beta} -
\frac{\lambda A_{\lambda}}{\sqrt{2}}v\cos{2\beta},\\
M_{25}^{2} & = & M_{52}^{2} = k\lambda v_{P}v\cos{2\beta},\\
M_{34}^{2} & = & M_{43}^{2} = -k\lambda v v_{P},\\
M_{35}^{2} & = & M_{53}^{2} = -k\lambda v v_{S}
+ \frac{\lambda A_{\lambda}}{\sqrt{2}}v,\\
M_{45}^{2} & = & M_{54}^{2} = 2k^{2}v_{S}v_{P} - \sqrt{2}kA_{k}v_{P}.
\ear
For the squared masses of the charged Higgs bosons one has
$m_{H^{\pm}}^{2} = m_{A}^{2} - \frac{\lambda^{2}v^{2}}{2} +
\frac{g^{2}v^{2}}{4}$.
From the expressions for the scalar mass matrices it can be easily
seen that, like in the minimal split SUSY, one can split the particle
spectrum by taking $\mu B$ parameter of order of the squared
splitting scale $m_{s}^{2}$. This means that the charged higgses, one 
scalar $H^{0}$ and one pseudoscalar $A^{0}$ from the Higgs sector
become heavy. It is straightforward to check that the low energy 
effective Lagrangian is obtained by the same substitution for the
Higgs doublets as in the minimal split SUSY \cite{Giudice:2004tc}
\be
\label{substit}
H_{u}\to H\sin{\beta},\;\;\;H_{d}\to\epsilon H^{*}\cos{\beta},
\ee
where $H$ is the Standard Model Higgs doublet. After the replacement
(\ref{substit}) and including interactions with fermions, we arrive at
the Lagrangian for light degrees of freedom
\[
\cal{L} = \cal{L}_{\rm{V}} + \cal{L}_{\rm{Y}}.
\]
It consists of two parts: the scalar potential
\begin{gather}
  -{\cal L_{\rm{V}}} =
  \frac{\bar{g}^{2}}{8}\cos^{2}{2\beta}\l(H^{\dagger}H\r)^{2} +
  |r + kN^{2} - \frac{\lambda}{2}\sin{2\beta}H^{\dagger}H|^{2} +   
  |\lambda N + \mu|^{2}H^{\dagger}H \nonumber\\
+ \l(-\frac{\lambda}{2}A_{\lambda}\sin{2\beta}NH^{\dagger}H -
\frac{\mu B}{2}\sin{2\beta}H^{\dagger}H + \frac{1}{3}kA_{k}N^{3} +
A_{r}N + h.c.\r)  \label{poten} \\
+ \l(m_{u}^{2}\sin^{2}{\beta} +
m_{d}^{2}\cos^{2}{\beta}\r)H^{\dagger}H + m_{N}^{2}|N|^{2} 
\nonumber
\end{gather}
and the Yukawa interactions
\be
 \label{yukava}
- {\cal L_{\rm{Y}}} = -\lambda N\tilde{H}_{u}\epsilon\tilde{H}_{d} - 
\lambda\sin{\beta}H^{T}\epsilon\l(\tilde{H}_{d}\tilde{n}\r) + 
\lambda\cos{\beta}\l(\tilde{n}\tilde{H}_{u}\r)H^{*} -
kN\tilde{n}\tilde{n} + h.c., 
\ee
where $\tilde{H}^{u}$ and $\tilde{H}^{d}$ are higgsinos and
$\tilde{n}$ is singlino, which is the fermionic component of the
singlet chiral superfield $\hat{N}$. In the above expressions we omitted   
terms with the Yukawa couplings of quarks to the Higgs bosons,  
gaugino interactions and mass terms, which are the same as the ones in
Ref.~\cite{Giudice:2004tc}. For this Lagrangian to describe the low
energy physics, all terms are to be of the order of electroweak scale. 
Inspecting the Higgs mass term in~(\ref{poten}) (see also
eq.~(\ref{fine})  below) and eqs.~(\ref{1.1}), (\ref{1.2}), one can
find that a strong cancellation in the combination
$m_{u}^{2}\tan{\beta} + m_{d}^{2}\cot{\beta} - 2B\mu$ has to take
place.

A special feature of our model, as compared to the minimal split SUSY,
is that after splitting there remain relatively light singlet fields:
complex scalar $N$ and Majorana fermion $\tilde{n}$. 

It is often assumed that NMSSM possesses ${\mathbb Z}_{3}$-symmetry,
which implies $\mu = r = A_{r} = B\mu = 0$. It was argued in
Ref.~\cite{Demidov:2004jx}, that the splitting of ${\mathbb
  Z}_{3}$-symmetric theory, which leads to the same particle content as
the one described above, results in the relations $\lambda \sim
\l(m_{ew}/m_{s}\r)^{2}$ and $\langle  N\rangle \sim 
m_{s}^{2}/m_{ew}$; therefore, the cubic term in the scalar 
effective potential behaves effectively like quadratic during the
electroweak phase transition. This means that EWPT cannot be strong
enough in that case and electroweak baryogenesis does not work. Thus
we do not impose the ${\mathbb Z}_{3}$-symmetry and take all
parameters in (\ref{superpotential}) and (\ref{bpotential}) to be
non-zero.

The Lagrangian (\ref{poten}), (\ref{yukava}) describes interactions at
the splitting scale $m_{s}$, which is taken below to be $10^{9}$~GeV,
if not stated otherwise. To obtain the low energy theory, the
couplings in (\ref{poten}), (\ref{yukava}) should be changed according
to the renormalization group equations (RGE). The set of RGE for
dimensionless couplings is presented in Appendix~A. 

\begin{figure}[hbt]
\begin{center}
\begin{picture}(500,100)(30,420)

\DashLine(60,470)(90,470){5}
\BCirc(110,470){20}
\DashLine(130,470)(160,490){5}
\DashLine(130,470)(160,450){5}
\put(60,475){{\small$N$}}
\put(105,494){{\small$H$}}
\put(105,443){{\small$H$}}
\put(152,492){{\small$N^{*}$}}
\put(152,443){{\small$N$}}

\DashLine(220,490)(250,470){5}
\DashLine(220,450)(250,470){5}
\BCirc(260,470){20}
\DashLine(280,470)(310,490){5}
\DashLine(280,470)(310,450){5}
\put(223,492){{\small$N$}}
\put(223,444){{\small$N$}}
\put(256,494){{\small$H$}}
\put(256,443){{\small$H$}}
\put(300,492){{\small$N$}}
\put(300,444){{\small$N$}}

\DashLine(370,490)(400,470){5}
\DashLine(370,450)(400,470){5}
\BCirc(410,470){20}
\DashLine(430,470)(460,490){5}
\DashLine(430,470)(460,450){5}
\put(373,492){{\small$N^{*}$}}
\put(373,444){{\small$N$}}
\put(407,494){{\small$H$}}
\put(407,443){{\small$H$}}
\put(450,492){{\small$N$}}
\put(450,444){{\small$N$}}

\end{picture}
\caption[ ]{\label{examples} Examples of the Feynman diagrams, which
  generate terms in the last line in eq.~(\ref{gener_poten}).}
\end{center}
\end{figure}
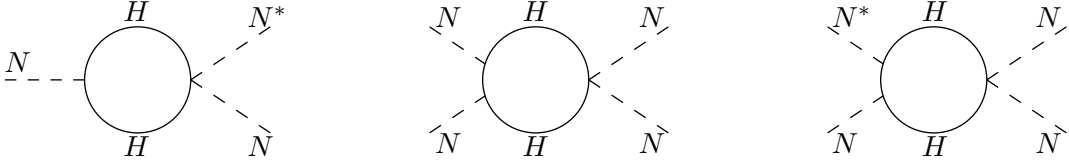

Below the splitting scale $m_{s}$, the theory is described by the
following Lagrangian
\begin{gather}
-{\cal L_{\rm{V}}} = -m^{2}H^{\dagger}H + 
\frac{\tilde{\lambda}}{2}\l(H^{\dagger}H\r)^{2} + 
i\tilde{A}_{1}H^{\dagger}H\l(N - N^{*}\r) + 
\tilde{A}_{2}H^{\dagger}H\l(N + N^{*}\r) +
\kappa_{1}|N|^{2}H^{\dagger}H \nonumber \\
  \label{gener_poten}
 - \kappa_{2}H^{\dagger}H\l(N^{2} + N^{*2}\r) + 
  \tilde{m}_{N}^{2}|N|^{2} + \lambda_{N}|N^{2}|^{2} + 
  \frac{1}{3}\tilde{A}_{k}\l(N^{3} + N^{*3}\r)
+ \tilde{A}_{r}\l(N + N^{*}\r) \\
+ \l(\frac{\tilde{m}^{2}}{2}N^{2} + \frac{1}{2}\tilde{A}_{3}N^{2}N^{*} +
\xi N^{4} + \frac{\eta}{6}N^{3}N^{*} + h.c.\r) \nonumber
\end{gather}
and
\begin{gather}
  - {\cal L_{\rm{Y}}} = \frac{M_{2}}{2}\tilde{W}^{a}\tilde{W}^{a} + 
  \frac{M_{1}}{2}\tilde{B}\tilde{B} +
  \l(\mu + \kappa N\r)\tilde{H}^{T}_{u}\epsilon\tilde{H}_{d} -
  kN\tilde{n}\tilde{n} \nonumber \\
+ H^{\dagger}\l(\frac{1}{\sqrt{2}}\tilde{g}_{u}\sigma^{a}\tilde{W}^{a}
+ \frac{1}{\sqrt{2}}\tilde{g}_{u}^{\prime}\tilde{B} -
\lambda_{u}\tilde{n}\r)\tilde{H}_{u}   \label{gener_yukava} \\
+ H^{T}\epsilon\l( 
-\frac{1}{\sqrt{2}}\tilde{g}_{d}\sigma^{a}\tilde{W}^{a} +
\frac{1}{\sqrt{2}}\tilde{g}^{\prime}_{d}\tilde{B} - 
\lambda_{d}\tilde{n}\r)\tilde{H}_{d} + h.c., \nonumber
\end{gather}
where we have also written explicitly the interactions between the
Higgs bosons, higgsinos and gauginos. The terms, analogous to ones
with couplings $\tilde{A}_{3}, \xi, \eta$ in the last line in
eq.~(\ref{gener_poten}), are absent in  eq.~(\ref{poten}), but they
are generated below $m_s$ by quantum corrections coming from one-loop
diagrams of the types presented in Fig.~\ref{examples}. 
Comparing the Lagrangians (\ref{poten}), (\ref{gener_poten}) and
(\ref{yukava}), (\ref{gener_yukava}), we read off the following
matching conditions between the coupling constants, which are valid at
the splitting scale $m_{s}$, 
\begin{gather}
\label{fine}
m^{2} = -\l(m_{u}^{2}\sin^{2}{\beta} + m_{d}^{2}\cos^{2}{\beta}
- \mu B\sin{2\beta} + |\mu|^{2} - \lambda r\sin{2\beta}\r), \\
\tilde{A}_{1} = \lambda\mu,\;\;\;\;
\tilde{A}_{2} = -\lambda A_{\lambda}\sin{\beta}\cos{\beta}, \nonumber\\
\kappa_{1} = \lambda^{2},\;\;\;\;
\kappa_{2} = -\lambda k\sin{\beta}\cos{\beta},\;\;\;\;
\lambda_{N} = k^{2}, \nonumber\\
\tilde{A}_{k} = kA_{k},\;\;\;
\tilde{m}^{2} = 2kr,\;\;\;
\tilde{A}_{r} = A_{r},\;\;\;
\tilde{m}_{N}^{2} = m_{N}^{2},
\label{scalar_couplings} \\
\label{tilde_lambda}
\tilde{\lambda} = \frac{\bar{g}^{2}}{4}\cos^{2}{2\beta} +
\frac{\lambda^{2}}{2}\sin^{2}{2\beta},\;\;\;
\kappa = \lambda, \\
\label{singlino_couplings}
\lambda_{u} = \lambda\cos{\beta},\;\;\;\;
\lambda_{d} = - \lambda\sin{\beta}, \\
\tilde{A}_{3} = \eta = \xi = 0 \label{1}
\end{gather}
and, like in the minimal split SUSY,
\be
\label{gaugino_couplings}
\tilde{g}_{u} = g\sin{\beta},\;\;\;
\tilde{g}_{d} = g\cos{\beta},\;\;\;
\tilde{g}_{u}^{\prime} = g^{\prime}\sin{\beta},\;\;\;
\tilde{g}_{d}^{\prime} = g^{\prime}\cos{\beta}.
\ee
The aforementioned cancellation in  (\ref{fine}) results in $m^{2}\sim
m_{ew}^{2}$. The matching conditions provide the initial values for
RGE. We use 2-loop RGE for gauge couplings and 1-loop RGE for other
dimensionless couplings, neglecting threshold effects. The coupling
constants, additional to the minimal split SUSY, begin to contribute
to the running of the gauge couplings only at 2-loop level and do not
spoil their unification. 

In the next section we investigate EWPT in this model and study the properties
of one-loop effective potential. Its zero temperature part
\cite{Coleman:1973jx} is written in $\overline{\rm DR}$ scheme as
follows, 
\be
\label{zero_temp}
V^{(1)} =
\sum_{i}(\pm)\frac{1}{64\pi^{2}}n_{i}m_{i}^{4}\left[\log{\frac{m_{i}^{2}}{q^{2}}}
  - \frac{3}{2}\right],
\ee
where the sum runs over all particle species of field-dependent mass
$m_{i}$ with $n_{i}$ degrees of freedom. Upper and lower signs
correspond to the boson and fermion cases, respectively. The
renormalization scale $q$ is chosen to be $100$~GeV. We include loop
contributions from top quark, gauge and Higgs boson, singlet
(pseudo)scalars. Let us single out the terms in $V_{tree}\equiv
-{\mathcal L}_{V}$, which are
quadratic in vev's $v,\;v_{S}$ and $v_{P}$, e.g.,
\be
V_{tree} = -\frac{m^{2}}{2}v^{2} + \frac{m_{S}^{2}}{2}\l(v_{S}^{2} + 
  v_{P}^{2}\r) + \frac{\tilde{m}^{2}}{2}\l(v_{S}^{2} - v_{P}^{2}\r) +
  V_{tree}^{>2},
\ee
where the last term denotes the rest of the tree level potential. 
Imposing the conditions 
\begin{gather}
\label{extrema1}
- m^{2} + \frac{1}{v}\frac{\der}{\der v}\l(V_{tree}^{>2} + V^{(1)}\r)
= 0\;, \\
\label{extrema2}
m_{S}^{2} + \tilde{m}^{2} + \frac{1}{v_{S}}\frac{\der}{\der
  v_{S}}\l(V_{tree}^{>2} + V^{(1)}\r) = 
m_{S}^{2} - \tilde{m}^{2} + \frac{1}{v_{P}}\frac{\der}{\der
  v_{P}}\l(V_{tree}^{>2} + V^{(1)}\r) =  
0\;,
\end{gather}
one ensures that minimum remains at $(v, v_{S}, v_{P})$.

Let us describe the part of the parameter space, relevant for our
study. There are three dimensionless parameters $\tan{\beta},
\;\lambda,\;k$, which {\it a priori} take any values compatible with
the weak coupling regime in both high-energy and low-energy theories. 
In addition, the low-energy spectrum has to be phenomenologically
viable. To reduce the parameter space of the model, in what follows we 
fix these dimensionless parameters at the splitting scale as 
(unless stated otherwise), 
\be
\label{set_param}
\tan{\beta}=10,\;\;\;
\lambda = 0.6,\;\;\;
k=-0.5.
\ee
There is nothing special in this set, except of the choice of rather
large value of $\tan{\beta}$. As it will be argued in 
Sec.~\ref{bau_section}, if $\tan{\beta}\sim 1$, produced amount of the
baryon asymmetry is zero within the semiclassical approximation
for CP-violating sources. Quantitatively, the numbers we obtain (the
exact amount of the baryon asymmetry, predictions for EDMs, etc.) are
stable with respect to small variations of these parameters about the
point (\ref{set_param}). 

The dimensionful parameters are taken at the electroweak scale. So, we
do not assume universal boundary conditions for soft supersymmetry
breaking terms. Below we vary some of the parameters keeping several
relations between them to simplify the analysis. For
$m^{2},\;\tilde{m}_{N}^{2}$ and $\tilde{m}^{2}$ we substitute the 
corresponding expressions in terms of $v,\;v_{S}$ and $v_{P}$, which
follow from eqs.~(\ref{extrema1}), (\ref{extrema2}). Since we take
$\mu$ to be imaginary and CP is violated in the Higgs sector, two
scalars and one pseudoscalar generally mix. We restrict our
considerations (the only reason is to reduce the number of free
parameters) to the case where mixing between the physical Higgs boson
and two other (pseudo)scalars is absent; in fact this can be achieved
by tuning trilinear constants $\tilde{A}_{1},\;\tilde{A}_{2}$. Nonzero
$\tilde{A}_{3}$ is generated by radiative corrections, hence
generically it is small and is not very important, 
so we set it equal to zero. We also choose $\tilde{A}_{r} = 0$ for
concreteness. The free dimensionful parameters we are left with are
vev's of singlet scalars $v_{S},\;v_{P}$, gaugino masses
$M_{1},\;M_{2},\;M_{3}$ and trilinear coupling constant
$\tilde{A}_{k}$. 

We use the RGE as follows. We take known values of the gauge and top
Yukawa couplings derived from the following relations
\footnote
{
The adopted value of $y_{t}$ corresponds to the top quark mass
$171.6$~GeV. This value is consistent with the latest preliminary 
combined results of CDF and D0~\cite{Brubaker:2006xn} $m_{t} = 171.4 
\pm 2.1$~GeV ($y_{t}\sim 0.94 - 0.97$).
}
: $\alpha_{s}(M_{Z})=0.117$, $y_{t}(m_{t})=0.95$, $M_{Z}=91.19$~GeV,
$M_{W}=80.42$~GeV. These four couplings run up to the scale
$\mu=m_{s}$ according to truncated RGE, in which we leave only
$y_{t}$, $g_{i}$, $i=1,2,3$. At the splitting scale we take the set of
other parameters in (\ref{gener_poten}) and (\ref{gener_yukava}),
obeying initial conditions (\ref{set_param}) and matching conditions
(\ref{tilde_lambda}-\ref{gaugino_couplings}). 
Then all couplings run down to the electroweak scale in accordance
with the  full set of RGE. Because of additional coupling constants in
RGE, the procedure described above does not give the correct low
energy pattern of gauge and top Yukawa couplings we started
with. Therefore, we tune top Yukawa coupling at the splitting scale
$m_{s}$ and check that obtained values of $g_{1}, g_{2}, g_{3}$
and $y_{t}$ at the electroweak scale are within experimental error bars. 
Also we 
check that all couplings remain in the perturbative regime up to the
GUT scale.

\begin{figure}[htb]
\centerline{
\includegraphics[width=0.55\columnwidth,height=0.4\columnwidth]{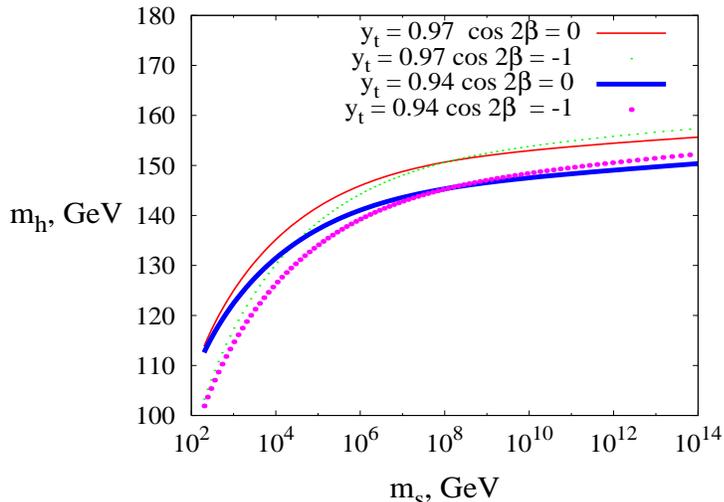}
}
\caption{\label{higgs_mass}
The dependence of the Higgs boson mass on the splitting scale $m_{s}$. 
Solid lines are for $\cos{2\beta} = 0$ case, while dotted ones are for 
$\cos{2\beta} = -1$. Top Yukawa coupling equals to $0.97$ for thin 
lines and $0.94$ for thick lines.
}
\end{figure}

In our restricted parameter space we use RGE to obtain the Higgs 
boson mass $m_{h}$. The dependence of $m_{h}$ on the splitting scale
$m_{s}$ is plotted in Fig.~\ref{higgs_mass}.
To obtain the plots shown in~Fig.~\ref{higgs_mass}, we use initial
conditions (\ref{set_param}) for $\lambda$ and $k$, 
and vary the value of $\beta$. In this analysis we take into account
the experimental uncertainties in the determination of the top quark
mass.

The values of the tree level Higgs boson mass are generally within the
same range as in the minimal split supersymmetry
\cite{Arkani-Hamed:2004fb,Arvanitaki:2004eu}. The upper bound
on $m_{h}$ is increased in comparison with MSSM. We have also found 
that the dependence of the Higgs boson mass $m_{h}$ on $k$ is within
$1\%$ in the whole perturbative range of $k$, while $m_{h}$ 
increases from $144.6$~GeV at $\lambda = 0.0$ to $160.0$~GeV at
$\lambda = 0.7$ (other parameters in this case are $k =
-0.5,\;\cos{2\beta} = 0,\;m_{s} = 10^{9}$~GeV).

Below we use the two reference points for the relevant parameters of
the model, which are presented in Table~\ref{points}.  
\begin{table}[htb]
{\renewcommand{\arraystretch}{1.2}
\begin{center}
\begin{tabular}{|c|c|c|c|c|c|c|}
\hline
& $v_{S}$ & $v_{P}$ & $\tilde{A}_{1}$ & $\tilde{A}_{2}$ &
$\tilde{A}_{k}$ & $\mu$ \\\hline
$(1)$ & $166$ & $310$ & $83.6$ & $-40.0$ & $-1.1$ & $69.6$ \\\hline
$(2)$ & $110$ & $-310$ & $-78.3$ & $-24.8$ & $-1.0$ & $-65.2$ \\\hline
\end{tabular}
\caption{\label{points} Two examples of parameters. All values
  are in GeV.
}
\end{center}
}
\end{table}
Both of them correspond to $m_{h} = 149$~GeV. 

\section{Electroweak phase transition}
\label{ewpt_section}

For successful electroweak baryogenesis the phase transition must be
strongly first order, so that the condition $v_{c}/T_{c}\gsim 1.1$
should be satisfied. Now we turn to the analysis of EWPT in our model
within the parameter space outlined above. The standard analytical way
to deal with this problem is based on the methods of finite
temperature field theory (see, e.g., Ref.~\cite{Quiros:1999jp} and
references therein).

Finite temperature one-loop effective potential for the Higgs and
singlet scalar fields reads as follows,
\be
\label{V_T}
V_{T}(v,v_{S},v_{P}) = V_{tree}(v,v_{S},v_{P}) +
V^{(1)}(v,v_{S},v_{P}) + V_{T}^{(1)}(v,v_{S},v_{P}).
\ee
Here the first and the second terms are the tree level part of the
potential (\ref{gener_poten}) and 1-loop contribution
(\ref{zero_temp}), respectively. The third term is the one-loop contribution
at finite temperature \cite{Dolan:1973qd},
\be
V_{T}^{(1)} = \sum_{i}f_{i}\l(m_{i}, T\r),
\ee
with
\be
f_{i}\l(m_{i}, T\r) = (\pm)\frac{T^{4}}{2\pi^{2}}
\int_{0}^{\infty} dx x^{2}\log{\l(1\mp e^{-\sqrt{x^{2} +
      (m_{i}/T)^{2}}}\r)},
\ee
where we use the same notations as in eq.~(\ref{zero_temp}). 
We found that in numerical calculations it is convenient to use
the following approximation 
\[
f_{B}\l(m, T\r) = -\frac{\pi^{2}T^{4}}{90}
+ \frac{m^{2}T^{2}}{24}
- \frac{m^{3}T}{12\pi}
- \frac{m^{4}}{64\pi^{2}}{\rm ln}\frac{m^{2}}{c_{B}T^{2}},\;\;{\rm for}\;
\frac{m}{T} < 2.2
\]
\[
f_{F}\l(m, T\r) = -\frac{7\pi^{2}T^{4}}{720} 
+ \frac{m^{2}T^{2}}{48}
+ \frac{m^{4}}{64\pi^{2}}{\rm ln}\frac{m^{2}}{c_{F}T^{2}},\;\;{\rm for}\;
\frac{m}{T} < 1.6,
\]
where ${\rm ln}\;c_{B} \approx 5.41$ and ${\rm ln}\;c_{F}
\approx 2.64$. The approximation 
\[
f_{B,F}\l(m, T\r) = - \l(\frac{m}{2\pi T}\r)^{3/2}T^{4}{\rm
e}^{-\frac{m}{T}}\l(1+ \frac{15T}{8m}\r),
\]
is valid for bosons at $m/T > 3.0$ and for fermions at $m/T > 2.2$.
We adopt linear interpolation between high- and low-temperature
regimes. 
We define the critical temperature $T_{c}$ as a temperature at which
the first bubbles of true vacuum begin to nucleate. It takes place
when $S_{3}\l(T_{c}\r)/T_{c}\sim 130-140$ \cite{Anderson:1991zb}
where $S_{3}\l(T\r)$ is the free energy of critical bubble. The
critical bubble is a saddle point of the free energy functional,
which for spherically symmetric configurations reads
\[
S_{3}\l(T\r) = 4\pi\int_{0}^{\infty}dr r^{2}\l[\l(\frac{dh}{dr}\r)^{2}
  + \l(\frac{dS}{dr}\r)^{2} + \l(\frac{dP}{dr}\r)^{2} + V_{T}\l(h, S, P\r)\r].
\]
Here the effective potential $V_T$ is defined by eq.~(\ref{V_T}), and 
shifted in such a way as to obey $V_{T}\l(0, S_{s}, P_{s}\r) = 0$,
where $S_{s}$ and $P_{s}$ are the values of the scalar and
pseudoscalar fields in the symmetric phase. The bubble profile
satisfies the equations of  motion 
\[
E_{h}\l(r\r) = \frac{d^{2}h}{dr^{2}} + \frac{2}{r}\frac{dh}{dr} -
\frac{\partial V_{T}}{\partial h} = 0,\;\;\;
E_{S}\l(r\r) = \frac{d^{2}S}{dr^{2}} + \frac{2}{r}\frac{dS}{dr} -
\frac{\partial V_{T}}{\partial S} = 0,\;\;\;
\]
\[
E_{P}\l(r\r) = \frac{d^{2}P}{dr^{2}} + \frac{2}{r}\frac{dP}{dr} -
\frac{\partial V_{T}}{\partial P} = 0
\]
and obeys the boundary conditions
\be
\label{BC_bounce}
\l(h(r), S(r), P(r)\r)|_{{r = \infty}} = \l(0, S_{s}, P_{s}\r),\;\;\;
\l(\frac{dh}{dr}, \frac{dS}{dr}, \frac{dP}{dr}\r)|_{r=0}
= \l(0, 0, 0\r).
\ee
To find the critical bubble profile numerically we use the method,
originally proposed in Ref.~\cite{Moreno:1998bq} for MSSM and further
developed in Ref.~\cite{John:1998ip}. Namely, we seek for the absolute
minimum of the following functional, 
\[
F\l[h(r),S(r),P(r)\r] = 4\pi\int_{0}^{\infty}dr r^{2}\l[E^{2}_{h}\l(r\r) +
  E^{2}_{S}\l(r\r) + E^{2}_{P}\l(r\r)\r],
\]
over the space of functions satisfying the boundary
conditions~(\ref{BC_bounce}). 

\begin{figure}[htb]
\centerline{
\includegraphics[width=0.7\columnwidth,height=0.45\columnwidth]{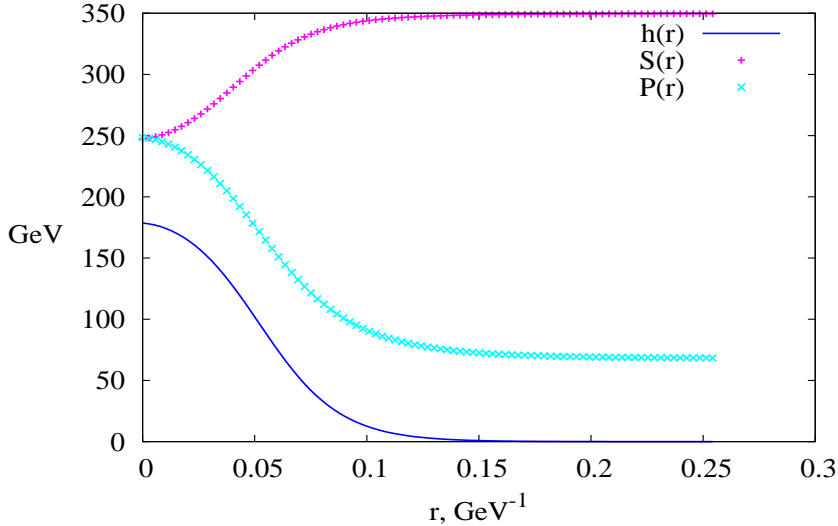}}
\caption{\label{bubble} 
  The critical bubble profile for the set $(1)$ in 
  Table~\ref{ewpt_points}. 
}
\end{figure}

We present the results of numerical analysis for parameters given in 
Table~1. In fact, these values were chosen to have the first order phase
transition. In Table~\ref{ewpt_points} we show the following
quantities: the critical temperature $T_{c}$, the critical values of
the Higgs, scalar and pseudoscalar fields, $v_{c},\;S_{c}$ and
$P_{c}$, respectively, and also the values of the scalar and
pseudoscalar fields, $S_{s}$ and $P_{s}$, in the symmetric phase at
the critical temperature. 
\begin{table}[htb]
\begin{center}
\begin{tabular}{|c|c|c|c|c|c|c|c|c|c|}
\hline
& $T_{c}$ & $v_{c}$ & $S_{c}$ & $P_{c}$ & $S_{s}$ & $P_{s}$ &
$S_{3}/T$ & $\delta_{T}\l(v_{c}\r)$ & $\delta_{T}\l(0\r)$\\\hline 
$(1)$ & $88$ & $218.4$ & $166.4$ & $308.4$ & $350$ & $68$ & $130 $ &
$0.13$ & $0.67$
\\\hline
$(2)$ & $96$ & $215$ & $110.8$ & $-307.6$ & $326.5$ & $-70.4$ &
$133$ & $0.18$ & $0.59$
\\\hline
\end{tabular}
\caption{\label{ewpt_points} Examples of parameters of the 
  strongly first order phase transition. All dimensionful values are
  in GeV. 
}
\end{center}
\end{table}
The critical bubble profile ({\it i.e.} the dependence of the scalar fields
on radial coordinate) for the set $(1)$ of the parameters in Table~1
is presented in Figure~\ref{bubble}.
The points in Table~2 correspond to the strongly ($v_{c}/T_{c} > 1$)
first order phase transition.

Let us comment briefly on the applicability of the perturbation theory
for the calculation of the effective potential $V_{T}$ near the critical 
temperature. Generally, the perturbation theory at finite temperature
is plagued by IR divergences coming from the gauge and scalar
sectors\footnote{
Fermions have no Matsubara zero modes, so their contributions
are finite in the infrared region.
}.
The IR divergences in the gauge sector invalidate the perturbative
calculations at $v = 0$. However, the small magnetic mass (of order of
$g^{2}T/\l(4\pi\r)$) of the gauge bosons is generated
nonperturbatively, and it serves the infrared cutoff for the
perturbation theory. It was shown in Ref.~\cite{Bodeker:1993kj} that
the effect of the magnetic mass renders the one-loop effective
potential a good approximation even near $v = 0$. This effect comes
from the interaction of gauge 
bosons and in our model it is the same as in SM. As
for scalar sector, the dimensionless expansion parameter $\delta_{T}$
for unresummed effective potential can be estimated as $\delta_{T}(v)
= \tilde{\delta}T^{2}/6 m^{2}_{T}(v)$ \cite{Arnold:1992rz}, where
$m_{T}(v)$ is the smallest thermal mass of scalar particle (we do not
show explicitly the dependence of $\delta_{T}$ on vev's of singlet
scalar fields) and $\tilde{\delta}$ is the corresponding coupling constant, 
which, for our parameter choice, is equal to $\tilde{\lambda}\sim 
0.36$ (cf. eq.~(\ref{tilde_lambda})). We have calculated the thermal
mass and the expansion parameter at the critical temperature $T_{c}$
for 
both minima of the effective potential $V_{T}$ and present the results
in the last two columns of Table~\ref{ewpt_points}. One concludes that
the perturbation theory is applicable for the points in 
Table~\ref{ewpt_points}. 

\section{The baryon asymmetry}
\label{bau_section}

In this section we estimate the amount of the baryon asymmetry
produced during the phase transition. There are several approaches to
this problem (see, e.g.,
\cite{Cline:1997vk}, \cite{Carena:1997ki}, \cite{Konstandin:2004gy}
and references therein). We have chosen to use the diffusion
approximation for the 
description of the particle densities in the presence of moving bubble
wall and WKB approximation for the CP-violating sources. We take into
account CP-violation coming from chargino sector only. For the sake of
simplicity, we ignore the singlino sector contribution to the
diffusion equations leaving this for future study. 

According to the semiclassical picture, particles and antiparticles
have different dispersion relations in CP-violating background of the 
bubble wall \cite{Joyce:1994fu}. Initially, the asymmetry in their
densities emerges in the chargino sector and then, due to
interactions and diffusion processes, it is transmitted into the
densities of other particle species including left-handed fermions
and, finally, to the baryon asymmetry. 

We derive the diffusion equations along the lines of
Ref.~\cite{Huet:1995sh}. We describe the behaviour of the $i$-th type
of particles by the currents $j_{i\mu} = \l(n_{i},\vec{j}_{i}\r)$,
where $n_{i}$ is the particle density and $\vec{j}_{i}$ is the
current density. In the diffusion approximation the current densities have 
the form $\vec{j}_{i} = - D_{i}\vec{\nabla}n_{i}$, where $D_{i}$ are
diffusion constants, so diffusion equations can be written in the
form
\be
\dot{n}_{i} - D_{i}{\bf\nabla}^{2}n_{i} = I_{i}\;.
\ee
Here the dot denotes the time derivative, $I_{i}$ are source terms to
be described below. To the linear order in chemical potentials $\mu_{i}$,
one has 
\be
n_{i} = g_{i}\int
\frac{d^{3}k}{(2\pi)^{3}}\l[f(E_{i},\mu_{i}) -
  f(E_{i},-\mu_{i})\r] = \frac{k_{i}(m_{i}/T)T^{2}}{6}\mu_{i},
\ee
where $g_{i}$ is the number of degrees of freedom, $f(E_{i},\mu_{i})$
is boson (or fermion) distribution function,  factor $k_{i}(m_{i}/T)$
is   equal to $2$ ($1$) 
for one boson (fermion)  massless degree of freedom  ($m_i=0$) and is 
exponentially suppressed in the limit $m_{i}/T \to \infty$. 

First, we neglect CP-violating reactions and weak sphalerons. The
CP-conserving part of the source terms can be written (again, to 
linear  order in chemical potentials) as follows, 
\be
I_{i} = \sum_{r}\Gamma^{r}_{i}\sum_{k}\mu_{k},
\ee
where the first sum runs over all reactions with the rate
$\Gamma^{r}_{i}$, in which the $i$-th type of particles participates, and
the second sum runs over all types of particle species participating
in the $r$-th reaction. As an example, for the process $A + 2B \to C +
3D$ with rate $\Gamma$ the corresponding term for particle $A$ is
$I_{A} \supset -\Gamma_A\l(\mu_{A} + 2\mu_{B} - \mu_{C} -
3\mu_{D}\r)$.  

We neglect all Yukawa interactions except for the top quark. Then,
leptons do not participate in the processes of
diffusion of the particle-antiparticle asymmetry. Moreover, strong
sphalerons, induced by the operator $\prod_{i = 1}^{3}
\bar{u}_{Ri}u_{Li}\bar{d}_{Ri}d_{Li}$, 
where index $i$ runs over three  
generations of quarks, are the only processes which produce light
quarks; therefore, one algebraically constrains the quark
densities as follows,
\be
n_{Q_{1}} = n_{Q_{2}} = - 2n_{u_{R}} = - 2n_{d_{R}} = - 2n_{s_{R}} = -
2n_{c_{R}} = - 2n_{B},
\ee
where $n_{Q_{i}} \equiv n_{u_{Li}} + n_{d_{Li}}$, ($i=1,2,3$) are
the densities of left doublets and $n_{B}\equiv n_{b_{R}}$. The
conservation of the baryon number, along with aforementioned
constraints, results in $n_{B} + n_{T} + n_{Q} = 0$, where we use
the standard notations $n_{T} \equiv n_{t_{R}}$ and $n_{Q} \equiv
n_{Q_{3}}$.

Since we neglect the singlet sector contribution, we only write
down\footnote{ 
We work in basis of the fields in the unbroken phase.
}
the source terms for the Higgs doublet $H^{\prime} = \l(H^{+},
H^{0}\r)^{T}$ (we use prime here to save the notation $H$ for later
use), 
higgsinos $\tilde{H}_{u} = \l(\tilde{H}_{u}^{+},
\tilde{H}_{u}^{0}\r)^{T}$, $\tilde{H}_{d} = \l(\tilde{H}_{d}^{0},
\tilde{H}_{d}^{-}\r)^{T}$ and  heavy quarks $Q = (t_{L}, b_{L})^{T}$,
$t_{R}$ and $b_{R}$ . The gauge interactions are assumed to be in
equilibrium, so the chemical potentials of the gauge bosons are equal
to zero. Therefore, the chemical potentials of the particles in
$SU(2)_{L}$ doublets are equal and, for our purposes, it is sufficient
to consider only the sums of densities $n_{Q} = n_{t_{L}} +
n_{b_{L}}$, $n_{H^{\prime}} = n_{H^{+}} + n_{H^{0}}$,
$n_{\tilde{H}_{u}} = n_{\tilde{H}_{u}^{+}} + n_{\tilde{H}_{u}^{0}}$
and $n_{\tilde{H}_{d}} = n_{\tilde{H}_{d}^{0}} +
n_{\tilde{H}_{d}^{-}}$. We also make usual assumptions, that the
chemical potentials of winos and bino are zero, being driven by the
high rate of helicity flipping interactions induced by their Majorana
mass terms. 

The following interactions are considered in the diffusion equations: 

(i)   strong sphalerons with the rate $\Gamma_{ss} = 
  6\kappa_{ss}\frac{8}{3}\alpha_{s}^{4}T$, where $\kappa_{ss}= O(1)$
  \cite{Huet:1995sh}, 

(ii)  top quark Yukawa interactions $y_{t}\bar{Q}tH$ with the rate 
  $\Gamma_{Y}$, 

(iii) interactions due to the term
  $\tilde{\mu}\tilde{H}_{u}\epsilon\tilde{H}_{d}$ with the rate
  $\Gamma_{\mu}$, 

(iv) top quark mass effects with the rate $\Gamma_{m}$,

(v) Higgs boson self interactions with the rate $\Gamma_{H}$,

(vi) Higgs-higgsino-gaugino interactions with the rates $\Gamma_{u} $
and $\Gamma_{d}$ for up and down higgsino doublets, respectively.

We neglect the curvature of the bubble wall and suppose that the main
part of the baryon asymmetry is produced when the bubble expands
already in the stationary regime. It means that near the wall all
space-time dependent quantities should depend only on the 
combination $z + v_{w}t$, where $v_{w}$ is the bubble wall
velocity and $z$ is a coordinate along the axis, 
perpendicular to the wall,
whose positive direction points to the broken phase. In particular,
$n_{i}\l(\vec{r}, t\r) = n_{i}\l(z + v_{w}t\r)$ and hence $\dot{n}_{i} =
v_{w}n_{i}$. 

With above assumptions one writes the diffusion equations as follows 
\begin{eqnarray}
v_{w}n_{Q}^{\prime} - D_{q}n_{Q}^{\prime\prime} &  = &  -\l(\mu_{Q} -
  \mu_{T} -   \mu_{H^{\prime}}\r)\Gamma_{Y} - \l(\mu_{Q} -
  \mu_{T}\r)\Gamma_{m} \\
& & -   6\l(2\mu_{Q} - \mu_{T} +
  9\mu_{B}\r)\Gamma_{ss}, \nonumber\\  
v_{w}n_{T}^{\prime} -  D_{q}n_{T}^{\prime\prime} & = &  \l(\mu_{Q} -
  \mu_{T} - \mu_{H^{\prime}}\r)\Gamma_{Y} + \l(\mu_{Q} -
  \mu_{T}\r)\Gamma_{m} \\
 & &+ 3\l(2\mu_{Q} - \mu_{T} +
  9\mu_{B}\r)\Gamma_{ss}, \nonumber\\  
v_{w}n_{H^{\prime}}^{\prime} -  D_{h}n_{H^{\prime}}^{\prime\prime} 
  & = &   -\l(\mu_{H^{\prime}} - \mu_{Q} + \mu_{T}\r)\Gamma_{Y} - 
  \l(\mu_{H^{\prime}} - \mu_{\tilde{H}_{u}}\r)\Gamma_{u} \label{49}\\
 & & - \l(\mu_{H^{\prime}} + \mu_{\tilde{H}_{d}}\r)\Gamma_{d} -
  \mu_{H^{\prime}}\Gamma_{H},  \nonumber\\  
v_{w}n_{\tilde{H}_{u}}^{\prime} -
  D_{h}n_{\tilde{H}_{u}}^{\prime\prime}  & = &  -
  \l(\mu_{\tilde{H}_{u}} - \mu_{H^{\prime}}\r)\Gamma_{u} - 
  \l(\mu_{\tilde{H}_{u}} + \mu_{\tilde{H}_{d}}\r)\Gamma_{\mu} +
  S_{u}, \label{50}\\ 
v_{w}n_{\tilde{H}_{d}}^{\prime} -
  D_{h}n_{\tilde{H}_{d}}^{\prime\prime} & = 
  & - \l(\mu_{\tilde{H}_{d}} + \mu_{H^{\prime}}\r)\Gamma_{d} -
  \l(\mu_{\tilde{H}_{u}} + \mu_{\tilde{H}_{d}}\r)\Gamma_{\mu} +
  S_{d},
\end{eqnarray}
where it is also assumed that the diffusion constants $D_{q}$ are the
same for all quarks and $D_{h}$ for all higgses and higgsinos. Here we
include CP-violating sources $S_{u}$ and $S_{d}$, which we describe
later. Neglecting moderate RG effects one estimates the ratio
$\Gamma_{u}/\Gamma_{d} = \tan^{2}{\beta}$.  
We consider the limit
of large $\tan{\beta}$, when $\Gamma_{u} \gg \Gamma_{d}$, and assume
the interactions corresponding  to the rate  $\Gamma_{u} $ to be in
equilibrium, which implies the constraint $\mu_{H^{\prime}} -
\mu_{\tilde{H}_{u}} = 0 $. Introducing the notation
$\Gamma_{\tilde{\mu}} \equiv \Gamma_{\mu} + \Gamma_{d}$ and excluding
the terms with $\Gamma_{u}$ from eqs.~(\ref{49}), (\ref{50}), one can
write down the following linear combinations
\begin{eqnarray}
I_{H^{\prime}} + I_{\tilde{H}_{u}}  & = & -\l(\mu_{H^{\prime}} -
\mu_{Q} + \mu_{T}\r)\Gamma_{Y} - \l(\mu_{H^{\prime}} +
\mu_{H_{d}}\r)\Gamma_{\tilde{\mu}} - \mu_{H^{\prime}}\Gamma_{H} +
S_{d},\\  
I_{\tilde{H}_{d}} & = & -\l(\mu_{H_{d}} +
\mu_{H^{\prime}}\r)\Gamma_{\tilde{\mu}} + S_{u}.
\end{eqnarray}
Upon defining the densities
\[
n_{h} = n_{H^{\prime}} + n_{\tilde{H}_{u}} + 
n_{\tilde{H}_{d}},\;\;
n_{H} = n_{H^{\prime}} + n_{\tilde{H}_{u}} -
n_{\tilde{H}_{d}},\;\; 
\]
we obtain the following set of diffusion equations 
\begin{eqnarray}
v_{w}n_{Q}^{\prime} & = & D_{q}n^{\prime\prime}_{Q} - \Gamma_{Y}\left[ 
  \frac{n_{Q}}{k_{Q}} - \frac{n_{T}}{k_{T}} - \frac{n_{H} +
  n_{h}}{k_{H_{1}}}\right] - \Gamma_{m}\left[\frac{n_{Q}}{k_{Q}} -
  \frac{n_{T}}{k_{T}}\right] \\
&& - 6\Gamma_{ss}\left[2\frac{n_{Q}}{k_{Q}}
  - \frac{n_{T}}{k_{T}} + 9\frac{n_{Q} + n_{T}}{k_{B}}\right],
  \nonumber \\
v_{w}n_{T}^{\prime} & = & D_{q}n_{T}^{\prime\prime} + \Gamma_{Y}\left[ 
  \frac{n_{Q}}{k_{Q}} - \frac{n_{T}}{k_{T}} - \frac{n_{H} +
  n_{h}}{k_{H_{1}}}\right] + \Gamma_{m}\left[\frac{n_{Q}}{k_{Q}} -
  \frac{n_{T}}{k_{T}}\right] \\
&& + 3\Gamma_{ss}\left[2\frac{n_{Q}}{k_{Q}}
  - \frac{n_{T}}{k_{T}} + 9\frac{n_{Q} + n_{T}}{k_{B}}\right],\nonumber\\
v_{w}n_{H}^{\prime} & = & D_{h}n_{H}^{\prime\prime} + \Gamma_{Y}\left[ 
  \frac{n_{Q}}{k_{Q}} - \frac{n_{T}}{k_{T}} - \frac{n_{H} +
  n_{h}}{k_{H_{1}}}\right] - \Gamma_{H}\frac{n_{H} + n_{h}}{k_{H_{1}}} 
  +  S_{u} - S_{d},\\ 
v_{w}n_{h}^{\prime} & = & D_{h}n_{h}^{\prime\prime} + \Gamma_{Y}\left[
  \frac{n_{Q}}{k_{Q}} - \frac{n_{T}}{k_{T}} - \frac{n_{H} +
  n_{h}}{k_{H_{1}}}\right] - 2\l[\frac{n_{H}}{k_{H}} +
  \frac{n_{h}}{k_{h}}\r]\Gamma_{\tilde{\mu}} \\
  && - \frac{n_{H} + n_{h}}{k_{H_{1}}}\Gamma_{H}  + S_{u} +
  S_{d}, \nonumber
\end{eqnarray}
where 
\[
k_{H_{1}} = 2(k_{H^{\prime}} + k_{\tilde{H}_{u}}),\;\;\;
k_{H} = \frac{2(k_{H^{\prime}} +
  k_{\tilde{H}_{u}})k_{\tilde{H}_{d}}}{k_{\tilde{H}_{d}} -
  k_{H^{\prime}} -  k_{\tilde{H}_{u}}},\;\;\;
k_{h} = \frac{2(k_{H^{\prime}} +
  k_{\tilde{H}_{u}})k_{\tilde{H}_{d}}}{k_{\tilde{H}_{d}} +
  k_{H^{\prime}} +  k_{\tilde{H}_{u}}}.\;\;
\]
We note in passing that in the case $\tan{\beta}\sim 1$, the assumption,
that $\Gamma_{u}$ and $\Gamma_{d}$ are in equilibrium leads to a
single diffusion equation for $n_{H}$, which is sourced only by the
difference $S_{u} - S_{d}$. As we will discuss later, in the
semiclassical limit $S_{u} = S_{d}$, hence no baryon asymmetry is 
produced, however, other approaches to the calculation of the
CP-violating sources, see, e.g.,
\cite{Carena:1997ki,Riotto:1998zb,Lee:2004we}) can lead to non-zero 
difference $S_{u} - S_{d}$.

Making standard assumption that strong sphaleron and Yukawa
interactions are in equilibrium and follow procedure described in
Ref.~\cite{Carena:1997ki}, we find the relations
\be
\label{QT}
n_{T} = - \frac{k_{T}\l(2k_{B} + 9k_{Q}\r)}{k_{H_{1}}\l(9k_{Q} +
  9k_{T} + k_{B}\r)}\l(n_{H} + n_{h}\r),\;\;
n_{Q} = \frac{k_{Q}\l(9k_{T} - k_{B}\r)}{k_{H_{1}}\l(9k_{Q} +
  9k_{T} + k_{B}\r)}\l(n_{H} + n_{h}\r)
\ee
and the equations for densities $n_{H}$ and $n_{h}$
\be
\label{eq_higgs}
{\bf \mathcal A}\l(
\begin{array}{c}
n_{H}^{\prime}\\
n_{h}^{\prime}
\end{array}
\r)
=
{\bf D}\l(
\begin{array}{c}
n_{H}^{\prime\prime}\\
n_{h}^{\prime\prime}
\end{array}
\r)
- {\bf \mathcal G}\l(
\begin{array}{c}
n_{H}\\
n_{h}
\end{array}
\r)
+
{\bf S},
\ee
where
\be
\begin{split}
{\bf\mathcal A} = 
\l(
\begin{array}{cc}
1 & \frac{F}{F + G} \\
\frac{F}{F + G} & 1
\end{array}
\r),\;\;
{\bf D} = 
\l(
\begin{array}{cc}
\bar{D}_{q} + \bar{D}_{h} & \bar{D}_{q}\\
\bar{D}_{q} & \bar{D}_{q} + \bar{D}_{h}
\end{array}
\r),\;\;\\
{\bf \mathcal G} =
\l(
\begin{array}{cc}
\bar{\Gamma}_{m} + \bar{\Gamma}_{h} & \bar{\Gamma}_{m} +
\bar{\Gamma}_{h}\\
\bar{\Gamma}_{m} + \bar{\Gamma}_{h} + 2\bar{\Gamma}_{\tilde{\mu}} & 
\bar{\Gamma}_{m} + \bar{\Gamma}_{h} +
2\frac{k_{H}}{k_{h}}\bar{\Gamma}_{\tilde{\mu}}  
\end{array}
\r),\;\;
{\bf S} = \l(
\begin{array}{c}
\frac{G}{G+F}\l(S_{u} + S_{d}\r)\\
\frac{G}{G+F}\l(S_{u} - S_{d}\r)
\end{array}
\r),
\end{split}
\ee
where $\bar{\Gamma}_{i} = \frac{G}{G + F}\Gamma_{i}$, $\bar{D}_{q} =
\frac{F}{F + G}D_{q}$, $\bar{D}_{h} = \frac{G}{F + G}D_{h}$ and
\[
F = 9k_{Q}k_{T} + k_{Q}k_{B} + 4k_{T}k_{B},\;\;
G = k_{H_{1}}\l(9k_{Q} + 9k_{T} + k_{B}\r).
\]
Using (\ref{QT}), one finds the density of left-handed fermions
\begin{multline}
\label{n_L}
n_{Left} = n_{Q_1} + n_{Q_2} + n_{Q_3} = 4n_{T} + 5n_{Q} \\ =
 \frac{5k_{Q}k_{B} +   8k_{T}k_{B} - 9k_{Q}k_{T}}{k_{H_{1}}\l(k_{B}
 +  9k_{Q} + 9k_{T}\r)}\l(n_{h} +   n_{H}\r) \equiv A\cdot
 (n_{h} + n_{H}). 
\end{multline}
For our set of statistical factors ($k_{Q} = 6,\;k_{T} = 3,\;k_{B} =
3$, $k_{H^{\prime}} = 4,\; k_{\tilde{H}_{u}} = k_{\tilde{H}_{d}} = 2$
in the massless case), which is the same as in SM (except for the higgsino
sector), the value of constant $A$ equals zero. This is well known
Standard Model suppression \cite{Giudice:1993bb}. As shown in
Ref.~\cite{Carena:2004ha} for similar situation, the corrections due
to subleading $O(1/\Gamma_{ss})$ effects are small, while radiative
corrections to the thermal masses of particles give a sizable
contribution mainly due to the top Yukawa coupling, 
\be
\label{n_L1}
A = \frac{3}{2k_{H_{1}}}\l(-\frac{3y_{t}^{2}}{8\pi^{2}}\r).
\ee
The space-time distribution of left-handed fermions acts as a source
for baryon asymmetry \cite{Carena:1997ki}, which obeys the following
equation
\be
v_{w}n_{B}^{\prime}(z) = -\theta(-z)\left[n_{F}\Gamma_{ws}n_{Left}(z) +
  {\cal R}n_{B}\r], 
\ee
where $\Gamma_{ws}$ is the weak sphaleron rate ($\Gamma_{ws} =
6\kappa_{ws}\alpha_{w}^{5}T$, where $\kappa_{ws} = 20\pm 2$
\cite{Moore:1999fs}), ${\cal R}$ is the  relaxation coefficient
\cite{Cline:1997vk,Shaposhnikov:1987tw} (${\cal R} =
\frac{5}{4}n_{F}\Gamma_{ws}$), and $n_{F} = 3$ is the number of 
families. Here theta-function implies that the weak sphalerons are
active only in the unbroken phase. The solution of this equation can
be obtained analytically, and the resulting baryon asymmetry takes the
form 
\be
\label{n_L2}
n_{B} =
      -\frac{n_{F}\Gamma_{ws}}{v_{w}}\int_{-\infty}^{0}dzn_{Left}(z)e^{z{\cal 
      R}/v_{w}}.
\ee

In the WKB approximation, the source terms\footnote{
Here we include CP-violating sources only for charginos. We make the
standard assumption that the contribution of neutralinos produce a
correction by a factor of order unity to the final result for the
baryon asymmetry (in MSSM this factor is about $1.5$). } $S_{u}$ and
$S_{d}$ are derived from chargino dispersion relations. The mass
matrix of charginos reads as follows,  
\be
\label{charg_matr}
M_{ch} = 
\l(
\begin{array}{cc}
M_{2} & \frac{1}{\sqrt{2}}\tilde{g}_{u}h(z) \\
\frac{1}{\sqrt{2}}\tilde{g}_{d}h(z) & \tilde{\mu}(z)
\end{array}
\r),
\ee
where we introduce the notation $\tilde{\mu}(z) = \mu + 
\kappa\l(S(z) + iP(z)\r)/\sqrt{2}$. The expressions for the sources 
(up to overall factor) come from Ref.~\cite{Huber:2000mg},  
where it has been shown that, in fact, the CP-violating sources for
higgsinos $\tilde{H}_{u}$ and $\tilde{H}_{d}$ are equal, so only the
equation for density $n_{h}$ is sourced. The
expressions for the sources read 
\begin{multline}
\label{CPsources}
S_{u} = S_{d} = \frac{T^{2}D_{h}v_{w}}{6\langle p_{z}^{2}/E\rangle_{0}}
\Big\{
\la A\ra\big[\theta^{\prime} -
\gamma^{\prime}\sin^{2}{a} +
\delta^{\prime}\sin^{2}{b} \big]^{\prime} +\\
\la B \ra
(m_{\tilde{h}}^{2})^{\prime}
\big[
\theta^{\prime} - \gamma^{\prime}\sin^{2}{a} +
\delta^{\prime}\sin^{2}{b}
\big]
 \Big\}^{\prime},
\end{multline}
where
\be
\la A\ra = \l\la
\frac{|p_{z}||m_{\tilde{h}}|^{2}}{2E^{2}}\r\ra_{+},\;\;\; 
\la B\ra = \l\la \frac{|p_{z}|(E^{2} -
  |m_{\tilde{h}}|^{2})}{2E^4}\r\ra_{+}, 
\ee
and $E^{2} = {\bf p}^{2} + m_{\tilde{h}}^{2}$, $p_{z}$ is the
component of the spatial momentum perpendicular to the wall, and
$m_{\tilde{h}}$ is the mass eigenstate of the matrix
(\ref{charg_matr}), which becomes pure higgsino in the unbroken
phase. The thermal averages $\la\;\ra_{+}$ and $\la\;\ra_{0}$ are
performed with fermion distribution function
\be
\la A\l(\vec{p}\r)\ra = \frac{\int
  d^{3}p\l(df/dE\r) A\l(\vec{p}\r)}{\int d^{3}p 
  \l(df/dE\r)|_{m=0}},\;\;\;\frac{df}{dE} =
-\beta\frac{{\rm e}^{\beta E}}{\l({\rm e}^{\beta E} + 1\r)^{2}}
\ee
for massive $(+)$ and massless $(0)$ cases, respectively. The other
quantities in eq.~(\ref{CPsources}) come from the diagonalization of 
chargino mass  matrix~\cite{Huber:2000mg},
\begin{gather}
\sin^{2} a = 2|A|^{2}/\Lambda(\Lambda + \Delta),\;\;\;
A = h\l(\tilde{g}_{u}M_{2}
+\tilde{g}_{d}\tilde{\mu}\r)/\sqrt{2},\;\;\; 
\Lambda = \l(\Delta^{2} + 4|A|^{2}\r)^{1/2},\\
\gamma = {\rm arg}A,\;\;\;
\gamma^{\prime}\sin^{2}{a} = 2{\rm
  Im}\l(A^{*}A^{\prime}\r)/\Lambda\l(\Lambda + \Delta\r),\;\;\;
\Delta = M_{2}^{2} - |\tilde{\mu}|^{2} + \l(\tilde{g}_{d}^{2} -
\tilde{g}_{u}^{2}\r)h^{2} 
\end{gather}
The expressions for $\sin^{2}{b}$, $\delta$ and 
$\delta^{\prime}\sin^{2}{b}$ are obtained from the ones for
$\sin^{2}{a}$, $\gamma$ and $\gamma^{\prime}\cos^{2}{a}$ by replacement 
$a\to b$, $\tilde{g}_{u}\to\tilde{g}_{d}$ and $\gamma\to-\delta$. 

To calculate CP-violating sources, we need the bubble wall
profile. The standard way to find it is to solve the equations of
motion, derived from the effective potential $V_{T}(h,S,P)$ at the 
critical temperature $T_{c}$ (for recent calculations of the bubble
wall 
profiles in (N)MSSM see, e.g., Refs.~\cite{John:1998ip}). For our
purposes we approximate its form by the standard kink-like solution
\begin{gather}
h(z) = \frac{1}{2}v_{c}\l(1 - \tanh{\left[\alpha\l(1 -
    \frac{2z}{L_{w}}\r)\right]}\r),\\
\l(
\begin{array}{c}
S(z) \\
P(z) \\
\end{array}
\r)
= 
\l(
\begin{array}{c}
S_{c} \\
P_{c} \\
\end{array}
\r)
-
\frac{1}{2}\l(
\begin{array}{c}
\Delta v_{S} \\
\Delta v_{P} \\
\end{array}
\r)
\l(1 + \tanh{\left[\alpha\l(1 -
    \frac{2z}{L_{w}}\r)\right]}\r),
\end{gather}
where $S_{c}$, $P_{c}$ are the critical values of the singlet scalar
fields,  $\Delta v_{S} = S_{c} - S_{s}$ and $\Delta v_{P} = P_{c} -
P_{s}$ (see also Table~\ref{ewpt_points}), $L_{w}$ is the bubble wall
width, which is taken to be the same for all scalar fields, and the
constant $\alpha$ is taken to be $1.5$ as in
Ref.~\cite{Carena:1997ki}. Here we have assumed that at $z >
L_{w}/(2\alpha)$ gauge symmetry is broken (the inner region of the
bubble) and at $z < 0$ symmetry is restored (outer region). 

For numerical calculations we take the diffusion constants to be
$D_{q}\sim 6/T$ for quarks and $D_{h}\sim 110/T$ for Higgs sector
\cite{Joyce:1994zn}; the width and the velocity of the bubble wall are
taken as $L_{w} = 8.0/T$ and $v_{w} = 0.1$, respectively.  The rates
\be
\Gamma_{H} = 0.0036T\;\theta\l(z - 0.5L_{w}\r),\;\;\;
\Gamma_{m} = 0.05T\;\theta\l(z - 0.5L_{w}\r)
\ee
are the same as in Ref.~\cite{Huber:2000mg} and $\Gamma_{\tilde{\mu}}
= 0.1T$ \cite{Carena:2002ss}.  

\begin{figure}[htb]
\centerline{
\includegraphics[width=0.7\columnwidth,height=0.45\columnwidth]{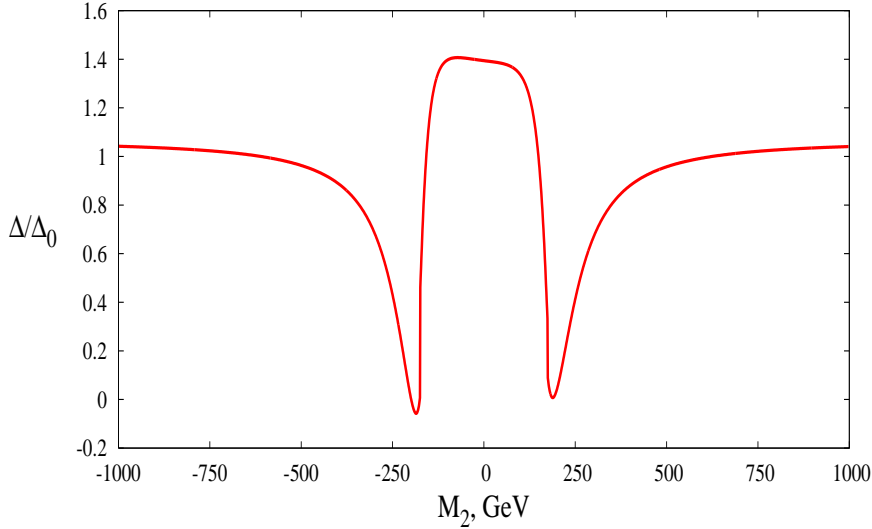}}
\caption{\label{M2} Plot of $\Delta/\Delta_{0}$ as a function of
  $M_{2}$   for the set $(2)$ of parameters presented in
  Table~\ref{ewpt_points}. 
}
\end{figure}

Below we present the results for the baryon-to-entropy ratio $\Delta =  
n_{B}/s$ with entropy density $s$ given by $s = 2\pi^{2}g_{eff}T^{3}/45,$
where $g_{eff}$ is the effective number of relativistic degrees of
freedom at the critical temperature. We plot the ratio $\Delta/ 
\Delta_{0}$, where $\Delta_{0} = 8.7\cdot 10^{-11}$, which corresponds to
$n_B/n_\gamma= 6.5\cdot 10^{-10}$ (cf. eq.~\eqref{BAU}).
In Fig.~\ref{M2}
we show the dependence of the baryon asymmetry on the gaugino mass 
$M_{2}$ for the set $(2)$ of the parameters in the Tables~1 and~2. We
found that in contrast to MSSM, the baryon 
asymmetry does not vanish  in the limit of infinite $M_{2}$, where 
one of the chargino species decouples. This fact may be attributed to
the strong $z$-dependence of the effective $\tilde{\mu}$ parameter. We
have checked numerically that in the limits $M_{2}\to +\infty$ and
$M_{2}\to -\infty$ the results for the baryon asymmetry coincide as it
should be, since in both cases charged gaugino (wino) is very heavy
and does not affect the low energy physics. Similarly, at $|M_{2}| <<
|\tilde{\mu}|$ the amount of baryon asymmetry is not very sensitive to
$M_{2}$ and is determined by $\tilde{\mu}(z)$ and off-diagonal
entries of the mass matrix (\ref{charg_matr}). In both regions
($M_{2}\to\infty$ and $M_{2}\to 0$) wino admixture in the relevant
higgsino-like mass eigenstate is small, so our approximation where we
neglect mixing in diffusion equations can be justified. The region of
large higgsino-wino mixing ($|M_{2}| \sim |\tilde{\mu}|\sim 200$~GeV
in Figure~\ref{M2}) should be considered more accurately if one wants 
to know the exact amount of baryon asymmetry produced in this case
(see \cite{Konstandin:2004gy} and references therein).  

To summarize this section, we demonstrated that the model is indeed
capable of producing sufficient amount of the baryon asymmetry. There
are certainly values of parameters for which the electroweak phase
phase transition is sufficiently strongly first order, and electroweak
baryogenesis is sufficiently efficient. In this analysis we restricted
ourselves to rather small range of parameters. It remains to be
understood how generic is successive electroweak baryogenesis in the
whole parameter space.

\section{Electric dipole moments}
\label{edm_section}

One of the consequences of the presence of additional sources
of CP-violation is their contribution to the electron and neutron
EDMs. Like in the minimal split SUSY model, there are three types of
contributions to EDM of a fermion (lepton or quark), related to the
exchange of $h_{0}\gamma$, $W^{+}W^{-}$ and $h_{0}Z$ bosons (see
Refs.
\cite{Arkani-Hamed:2004yi}, \cite{Chang:2005ac},
\cite{Deshpande:2005gi}, \cite{Giudice:2005rz}). 
The corresponding two-loop Feynman diagrams for SM fermion $f$
(charged lepton or quark) are given in Fig.~\ref{EDM_diag}.
\begin{tiny}
\begin{figure}[hbt]
\begin{center}
\begin{picture}(450,100)(0,350)

{\ArrowLine(20,370)(120,370)}
\DashLine(40,370)(55,400){5}
\Photon(100,370)(85,400){3}{6}
\BCirc(70,400){15}
\Photon(70,415)(70,440){3}{6}
\put(5,366){{$f$}}
\put(130,366){{$f$}}
\put(100,385){\mbox{{$\gamma$}}}
\put(77,425){{$\gamma$}}
\put(65,400){{$\chi_{j}^{\pm}$}}
\put(45,440){{$\mathbf{d^{h\gamma}}$}}
{\put(35,385){$h^{0}$}}

\ArrowLine(180,370)(280,370)
\Gluon(200,370)(215,400){-3}{6}
\Gluon(260,370)(245,400){-3}{6}
\Photon(230,415)(230,440){3}{6}
\BCirc(230,400){15}
\put(170,366){{$f$}}
\put(288,366){{$f$}}
\put(185,385){{$W^{\pm}$}}
\put(260,385){{$W^{\pm}$}}
\put(227,408){{$\chi_{j}^{\pm}$}}
\put(227,391){{$\chi_{i}^{0}$}}
\put(237,425){{$\gamma$}}
\put(198,437){{$\mathbf{d_{WW}}$}}

\ArrowLine(340,370)(440,370)
\Gluon(360,370)(375,400){-3}{6}
\DashLine(420,370)(405,400){5}
\Photon(390,415)(390,440){3}{6}
\BCirc(390,400){15}
\put(330,366){{$f$}}
\put(448,366){{$f$}}
\put(348,385){{$Z^{0}$}}
\put(418,385){{$h^{0}$}}
\put(387,400){{$\chi_{j}^{\pm}$}}
\put(397,425){{$\gamma$}}
\put(365,437){{$\mathbf{d_{Z\gamma}}$}}

\end{picture} \\
\caption[ ]{\label{EDM_diag} Feynman diagrams contributing to the
  fermion EDM in split SUSY.}
\end{center}
\end{figure}
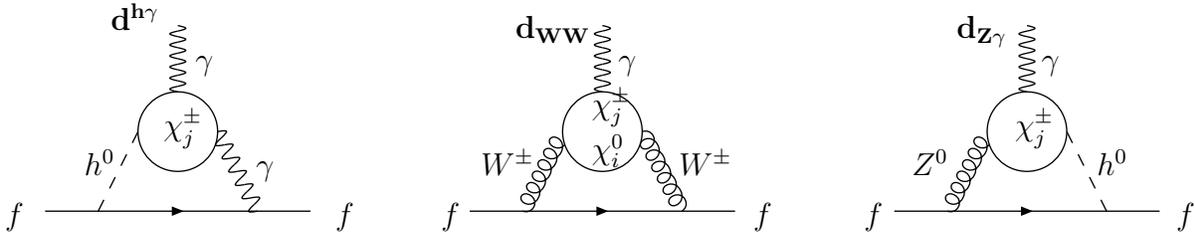
\end{tiny}
It was shown in Ref.~\cite{Giudice:2005rz} that the third contribution 
is suppressed in the case of electron by a factor ($1 -
4\sin^{2}{\theta_{W}}$), where $\theta_{W}$ is the weak mixing angle,
while it is important for the neutron EDM. In the minimal split SUSY
model there is only one CP-violating phase, associated with the
invariant $\phi_{1} = {\rm arg}(\tilde{g_{u}}^{*}\tilde{g_{d}}^{*}
M_{2} \tilde{\mu})$ where $\tilde{\mu} = \mu + \kappa\l(v_{S} +
iv_{P}\r)/\sqrt{2}$.  In the class of models presented here, among
parameters directly governing the couplings contributing to the
diagrams in Fig.~\ref{EDM_diag}, two additional independent phases
appear, which correspond to the phase invariants $\phi_{2} = {\rm
arg}\l(\kappa  
k^{*}\lambda_{u}\lambda_{d}(\tilde{\mu}^{*})^{-2}\r)$ and  
$\phi_{3} = {\rm arg}\l(\lambda_{u}\lambda_{d}^{*}
\tilde{g}_{u}^{*}\tilde{g}_{d}\r)$. 

After diagonalization of chargino and neutralino mass matrices,
the relevant part of the interaction Lagrangian becomes
\begin{gather}
{- \cal L_{\rm{int}}} =
\frac{g}{\cos_{\theta_{W}}}\bar{\chi}_{i}^{+}\gamma^{\mu}\l(G^{R}_{ij}P_{R} +
G_{ij}^{L}P_{L}\r)\chi_{j}^{+}Z_{\mu}\\
+ \left[g\bar{\chi}^{+}_{i}\gamma^{\mu}\l(C_{ij}^{R}P_{R} +
C_{ij}^{L}P_{L}\r)\chi_{j}^{0}W_{\mu}^{+} +
\frac{g}{\sqrt{2}}\bar{\chi}_{i}^{+}\l(D_{ij}^{R}P_{R} + 
D_{ij}^{L}P_{L}\r)\chi_{j}^{+}h + h.c.\right], \nonumber
\end{gather}
where the matrices are
\begin{gather}
  G_{ij}^{L} = V_{i1}c_{W^{+}}V^{\dagger}_{1j} +
  V_{i2}c^{h_{u}^{+}}V^{\dagger}_{2j},\;\;
  -G_{ij}^{R*} = U_{i1}c_{W^{-}}U^{\dagger}_{1j} +
  U_{i2}c_{h_{d}^{-}}U^{\dagger}_{2j}, \\
gD_{ij}^{R} = \tilde{g}^{*}_{u}V_{i2}U_{j1} +
\tilde{g}^{*}_{d}V_{i1}U_{j2},\;\;
gD^{L} = \l(gD^{R}\r)^{\dagger},
\end{gather}
with $i,j = 1,2$ and $c_{W^{\pm}} = \pm\cos^{2}{\theta_{W}}$,
$c_{h_{u}^{+},h_{d}^{-}} = \pm(1/2 - \sin^{2}{\theta_{W}})$, 
\be
C_{ij}^{L} = -V_{i1}N^{*}_{j2} + 
\frac{1}{\sqrt{2}}V_{i2}N^{*}_{j4},\;\;
C_{ij}^{R} = -U^{*}_{i1}N_{j2} -
\frac{1}{\sqrt{2}}U^{*}_{i2}N_{j3},\;\;
\ee
with $i = 1,2$ and $j = 1,\dots,5$. Here the unitary matrices $V$, $U$ 
and $N$ diagonalize the mass matrix of charginos (\ref{charg_matr})
and the mass matrix of neutralinos 
\be
\label{mass_neu}
M_{n} = 
  \l(
  \begin{array}{ccccc}
    M_{1} & 0 & -\frac{1}{2}\tilde{g}^{\prime}_{d}v &
    \frac{1}{2}\tilde{g}^{\prime}_{u}v & 0 \\
    0  &  M_{2} & \frac{1}{2}\tilde{g}_{d}v &
    \frac{1}{2}\tilde{g}_{u}v & 0 \\
    -\frac{1}{2}\tilde{g}^{\prime}_{d}v & \frac{1}{2}\tilde{g}_{d}v &
    0 & -\tilde{\mu} & \frac{1}{\sqrt{2}}\lambda_{d}v \\
    \frac{1}{2}\tilde{g}^{\prime}_{u}v & \frac{1}{2}\tilde{g}_{u}v
    & -\tilde{\mu} & 0 & \frac{1}{\sqrt{2}}\lambda_{u}v \\
    0 & 0 & \frac{1}{\sqrt{2}}\lambda_{d}v &
    \frac{1}{\sqrt{2}}\lambda_{u}v & -\sqrt{2}k(v_{S} + iv_{P})
  \end{array}
  \r)
\ee
so that $U^{*}{\cal M}_{C}V^{\dagger} =
{\rm diag}(m_{\chi^{+}_{1}}, m_{\chi^{+}_{2}})$ and
$N^{*}{\cal M}N^{\dagger} ={\rm diag}(m_{\chi^{0}_{1}},
m_{\chi^{0}_{2}}, m_{\chi^{0}_{3}}, 
m_{\chi^{0}_{4}}, m_{\chi^{0}_{5}})$. 
The EDM of electron or light quark has the form 
\[
d_{f} = d_{f}^{h\gamma} + d_{f}^{hZ} + d_{f}^{WW},
\]
where expressions for the corresponding contributions can be found in
Ref.~\cite{Giudice:2005rz}. We modified them appropriately in our
case, 
because of different number of neutralinos. The neutron dipole moment
can be expressed in terms of quark EDMs by the relation
\cite{Pospelov:1999ha,Pospelov:2000bw} 
\be
d_{n} = (1\pm 0.5)\left[\frac{f_{\pi}^{2}m_{\pi}^{2}}{(m_{u} +
    m_{d})(225{\rm MeV})^{3}}\right]\l(\frac{4}{3}d_{d} -
    \frac{1}{3}d_{u}\r),
\ee
where $f_{\pi} = 92$~MeV and $m_{u}/m_{d} = 0.55$.

\begin{figure}[htb]
\begin{tabular}{ll}
\includegraphics[angle=0,width=0.45\columnwidth]{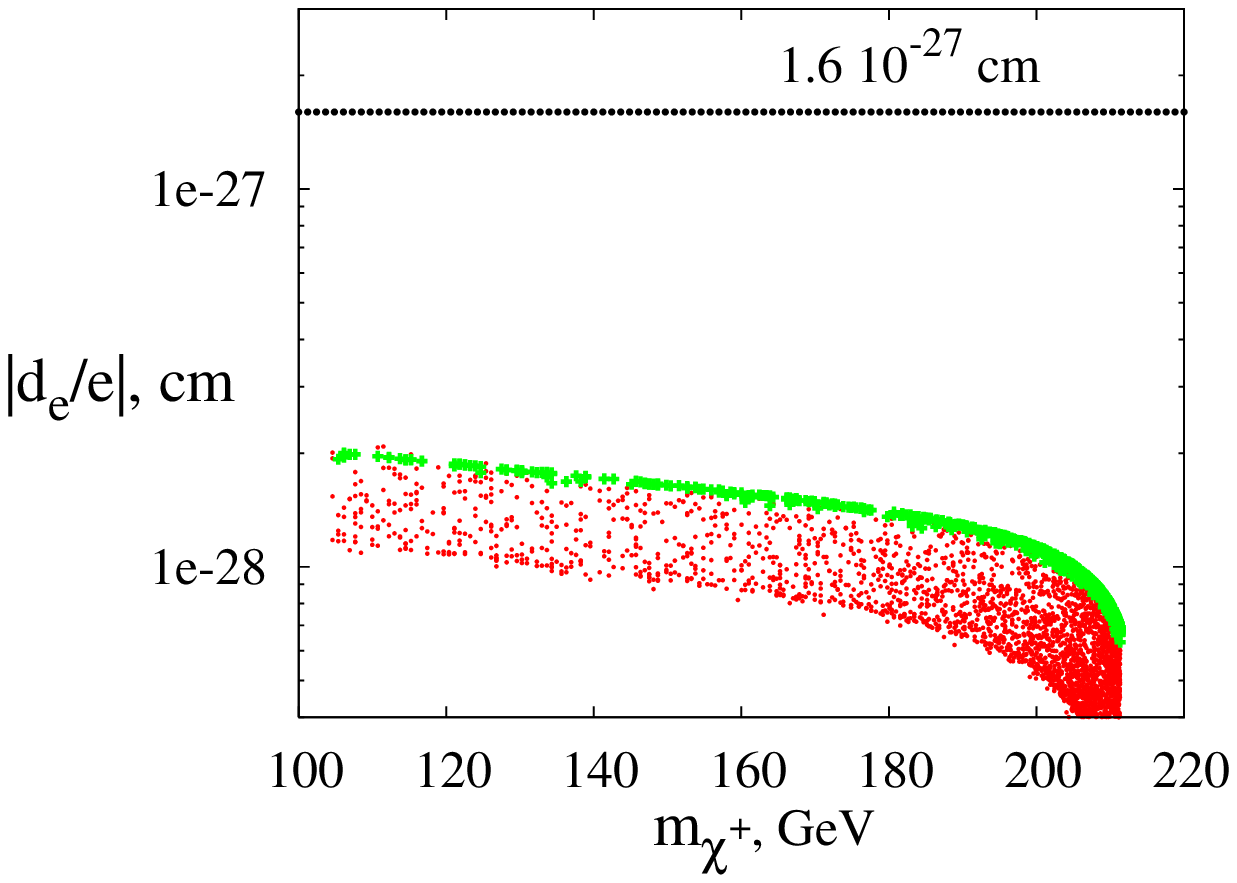} 
&
\includegraphics[angle=0,width=0.45\columnwidth]{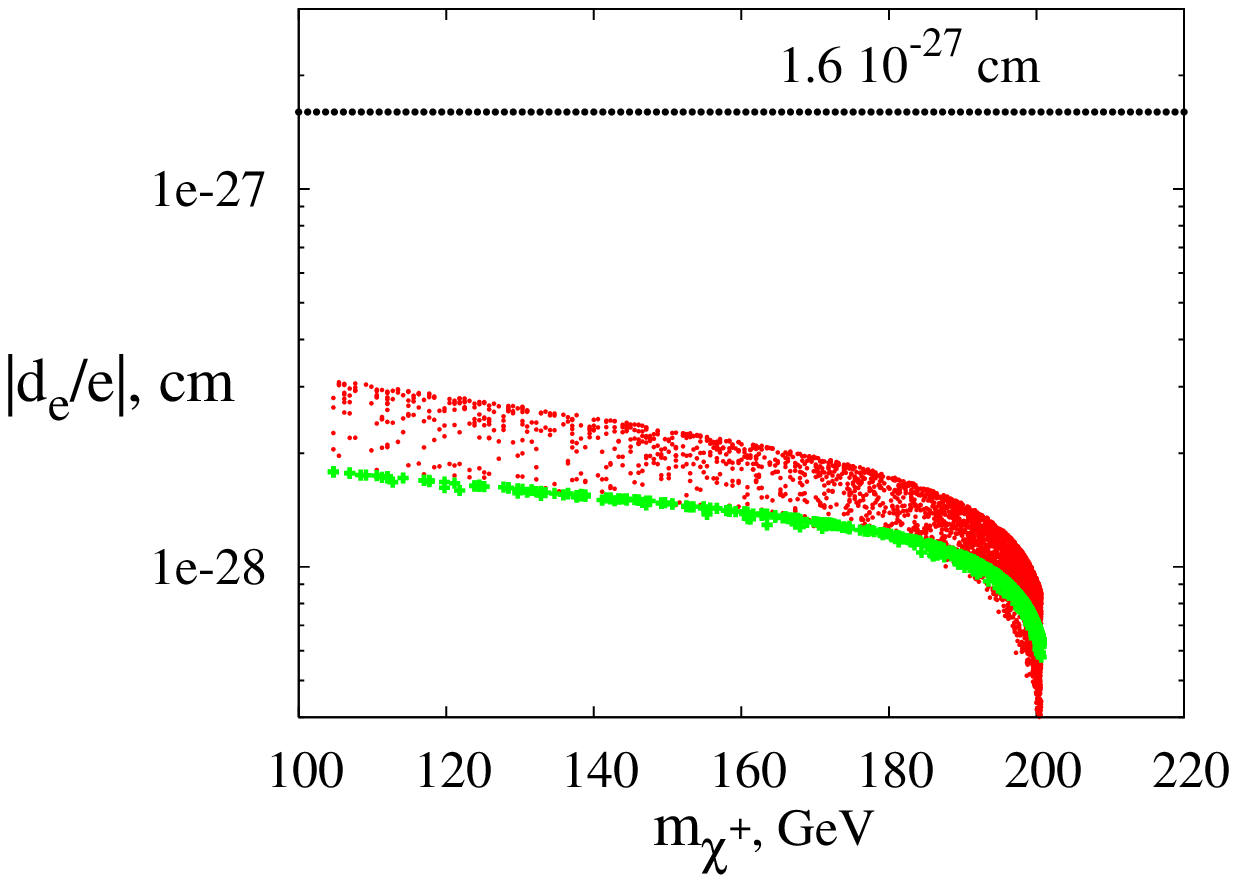} 
\end{tabular}
\caption{\label{e_edm} The EDM of electron as a function of the mass
  of the lightest chargino $m_{\chi^{+}}$ for set~1 (left) and set~2
  (right) of parameters in Table~1. Horizontal   dotted line
  represents the   experimental bound $|d_{e}| < 1.6\cdot 
  10^{-27} {\rm e\;cm}$. Green (gray) points represent the values of
  EDM for  real $k$. 
}
\end{figure}

\begin{figure}[htb]
\begin{tabular}{ll}
\includegraphics[angle=0,width=0.45\columnwidth]{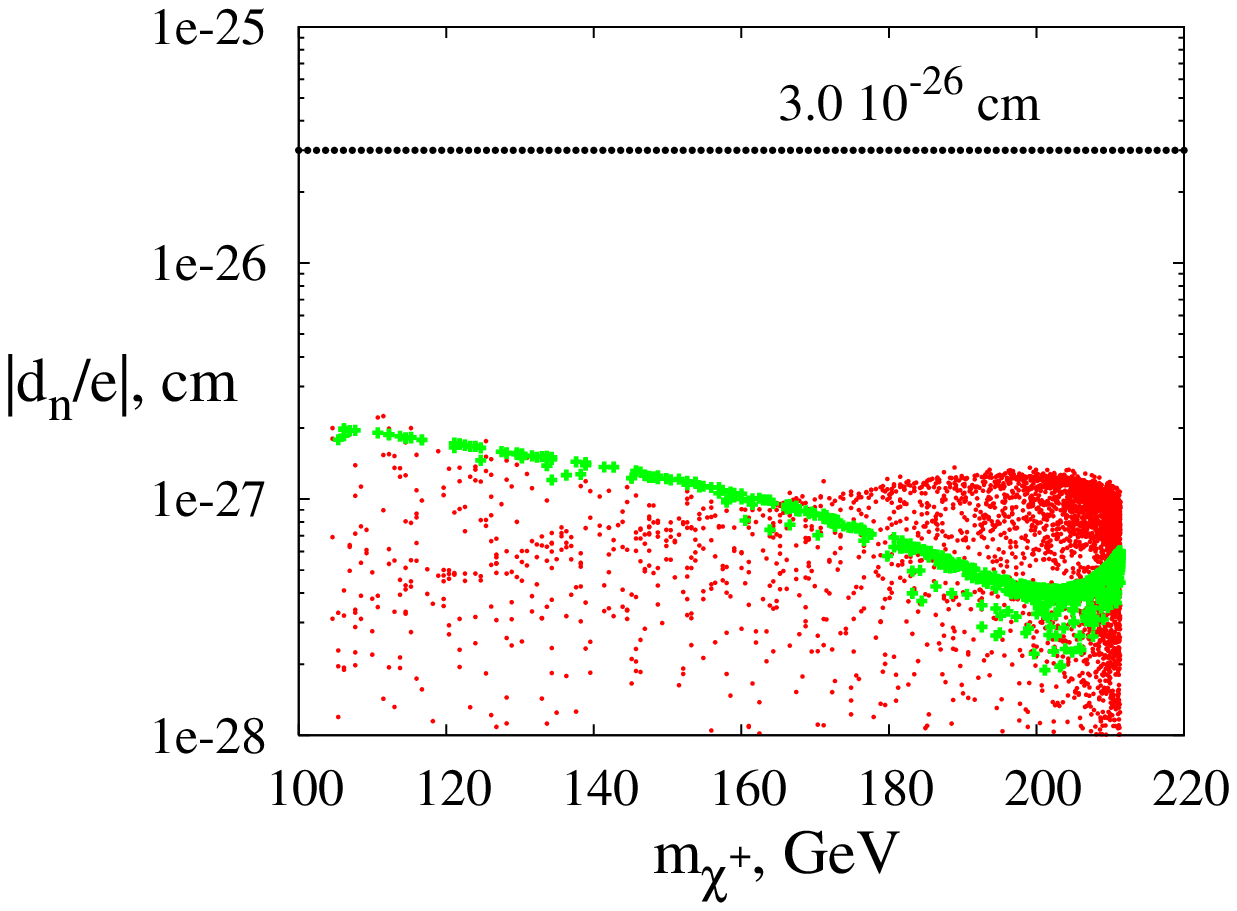} 
&
\includegraphics[angle=0,width=0.45\columnwidth]{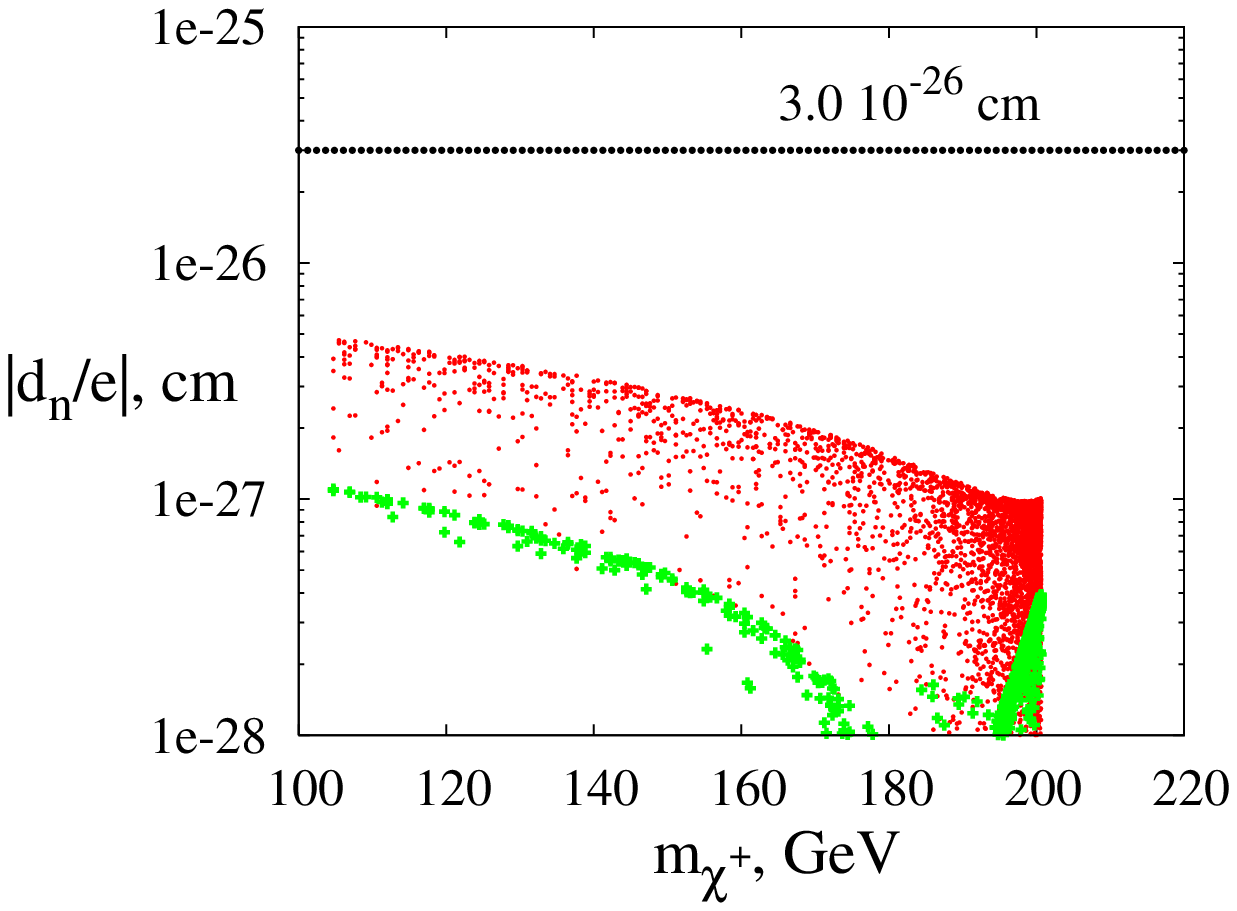} 
\end{tabular}
\caption{\label{n_edm} The EDM of neutron as a function of the
  mass of the lightest chargino $m_{\chi^{+}}$ for the same sets of
  parameters as in Fig.~\ref{e_edm}. Horizontal
  line represents the   experimental bound $|d_{n}| < 3.0\cdot
  10^{-26} {\rm e\;cm}$. Green (grey) points represent the values of
  EDM for real $k$. 
}
\end{figure}

For numerical calculations we use two sets of parameters (see
Table~1) and randomly scan over the 
following parameter space, $0 < M_{1}, M_{2} < 1000$~GeV. Also we take
the coupling $k$ to be complex, $k = |k|e^{i\phi_{k}}$, ($0 < \phi_{k}
< \pi$) to include the contribution of the phase invariant $\phi_{2}$.
The results for the electron and neutron EDMs as functions of the mass
of the lightest chargino are presented in Figs.~\ref{e_edm} and
\ref{n_edm}, respectively. We have taken into account the experimental
bound on the mass of the lightest chargino $m_{\chi^{+}} > 104$~GeV
\cite{Abbiendi:2003sc}. 

Horizontal lines show the present experimental limit on the EDM
of electron $|d_{e}| < 1.6\cdot 10^{-27}$~e~cm at
$90$\%~CL~\cite{Regan:2002ta} and neutron $|d_{n}| < 3.0\cdot   
10^{-26}$~e~cm at $90$\%~CL~\cite{Baker:2006ts}. One observes that 
generally for the cosmologically favorable models, the predictions for EDMs
are within one-two orders of magnitude below the present experimental
bound. Hence, our solution of the baryon asymmetry problem can be
indirectly probed by the future experiments aimed at EDM
searches. 

\section{Dark matter candidates}
\label{dark_matt}
In this section we explore dark matter in the model. In the minimal 
split SUSY this issue has been already investigated in
Refs.~\cite{Giudice:2004tc}, \cite{Arkani-Hamed:2004yi}, 
\cite{Pierce:2004mk}. In the first place, let us note that a
generalization of R-parity can be introduced in the nonminimal split
SUSY: with respect to this R-parity all new fermionic fields are odd,
while 
new (pseudo)scalars are even. Hence the lightest new fermion is the 
lightest superpartner and it is stable. 

The viable dark matter particles should be electrically neutral, so 
the best candidates in our model are neutralinos. We define the LSP 
neutralino state as 
\be
\chi = N_{51}\tilde{B} + N_{52}\tilde{W} + N_{53}\tilde{H}_{u}
+ N_{54}\tilde{H}_{d} + N_{55}\tilde{n}.
\ee
We estimate the neutralino relic abundance by using the standard
methods \cite{Gondolo:1990dk}. First we calculate the freeze-out
temperature $T_{F}$,
\be
x_{F} = {\rm log}
\l(\frac{m_{\chi}}{2\pi^{3}}\sqrt{\frac{45}{2g_{*}G_{N}x_{F}}}
\langle \sigma v\rangle_{\rm M\o l} \r),
\ee
where $x_{F}=m_{\chi}/T_{F}$, $g_{*}$ is the effective number of 
degrees of freedom at freeze-out $g^{1/2}_{*}\sim 9$; the
thermally-averaged product of neutralino annihilation cross section
$\sigma$ and relative velocity $v$ of neutralinos (the M{\o}ller cross
section $\langle \sigma v\rangle_{\rm M\o l}$) is 
\be
\langle \sigma v\rangle_{\rm M\o l} = 
\frac{1}{8m_{\chi}^{4}TK_{2}^{2}(m_{\chi}/T)}
\int_{4m_{\chi}^{2}}^{\infty}ds\sigma(s)(s - 4m_{\chi}^{2}) 
\sqrt{s}K_{1}\l(\frac{\sqrt{s}}{T}\r).
\ee
Here $K_{1}$, $K_{2}$ are the modified Bessel functions, $s$ is the 
Mandelstam variable. The relic abundance of the lightest neutralino is
then given by 
\be
\Omega_{\chi} h^{2} = \frac{(1.07\times 10^{9} {\rm
    GeV}^{-1})}{M_{Pl}} \l(\int_{x_{F}}^{\infty}dx 
\frac{\langle\sigma v\rangle_{\rm M\o
    l}(x)}{x^{2}}g_{*}^{1/2}\r)^{-1}.
\ee
To calculate the neutralino abundance, we modify the formulas for the 
annihilation cross sections presented in Ref.~\cite{Nihei:2002ij}. The
changes concern the annihilation of neutralinos into the Higgs bosons
and are due to new Feynman diagrams with singlet scalar and
pseudoscalar exchanges in s-channel. The corresponding
modifications are presented in Appendix~B. The total number of
neutralinos has been also changed according to our model.  

\begin{figure}[htb]
\begin{tabular}{ll}
\includegraphics[angle=0,width=0.45\columnwidth]{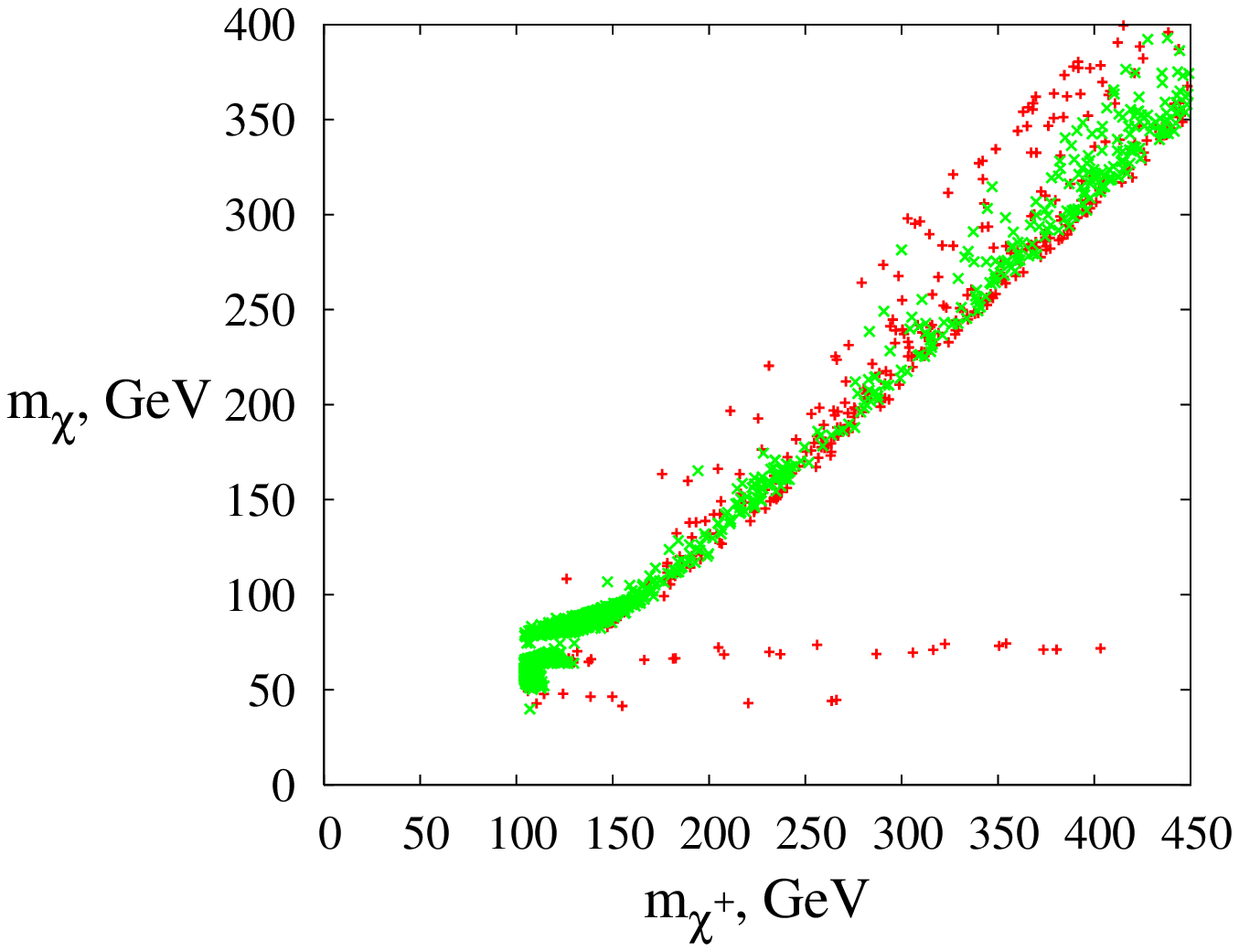} &
\includegraphics[angle=0,width=0.45\columnwidth]{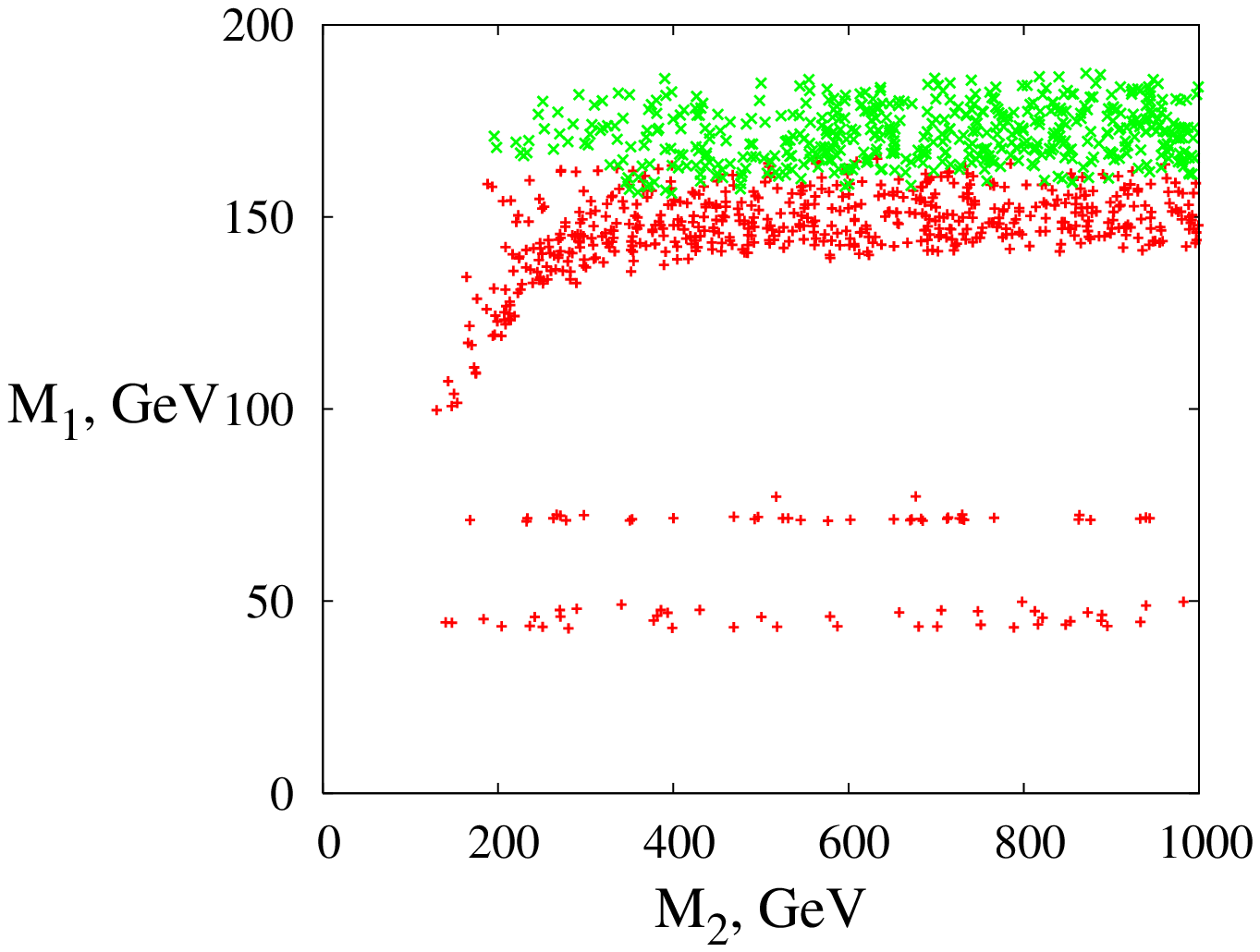}
\end{tabular}
\caption{\label{fig_dark_matt}
Points in ($m_{\chi^{+}}$, $m_{\chi}$) plane, {\it i.e.} the lightest
chargino and the lightest neutralino (left) and in ($M_{2},\;M_{1}$)
plane (right), which correspond to models with the correct relic
abundance of singlino ($|N_{55}|>0.5$, green (light grey) oblique
crosses) and non-singlino  (red / dark grey crosses) dark matter.  
}
\end{figure}

Adopting constraints discussed in Sec.~\ref{model_section}, we first
scan uniformly over the following parameter space: $|M_{1}|$, $|M_{2}|
< 1000$~GeV, $|v_{P}| < 2000$~GeV, with $v_{S}$ being in the region of
correct electroweak vacuum ({\it i.e.} the electroweak breaking
minimum is the global minimum of the potential) and squared mass
matrix of the scalar fields being diagonal. The numerical results are
presented in the left plot in Fig.~\ref{fig_dark_matt}, where we show
the region in $(m_{\chi^{+}},\; m_{\chi}$) plane favored by WMAP data:
each point  corresponds to a model in which neutralino abundance is
within the range $0.094<\Omega_{DM}h^{2}<0.129$ \cite{Bennett:2003bz}.
To check that the baryon asymmetry and dark matter problems can be
solved 
simultaneously, we also scan uniformly over the region in the
parameter space preferred by electroweak baryogenesis (cf.
Sec.~\ref{bau_section}); namely, we use $|M_{1}|,\;|M_{2}| <
1000$~GeV, with other parameters corresponding to the set $(2)$ in 
Tables~1 and~2. Points in ($M_{1}, \;M_{2}$) plane, which correspond
to correct neutralino abundance, are  shown on the right plot in
Fig.~\ref{fig_dark_matt}.   

On both plots green (light grey) crosses correspond to the dark matter
particles which have considerable admixture of singlino ($|N_{55}| >
0.5$), 
while the red (dark grey) crosses correspond to the mostly bino LSP. 
The annihilation of DM particles (bino as well as singlino) 
with masses $m_{\chi} \sim 0.5 M_{h} \sim 75~$GeV or $m_{\chi} \sim
0.5 M_{Z}$ proceeds resonantly via Higgs or $Z^{0}$-boson exchange,
respectively. On the right plot, this light neutralino corresponds to
the red (dark grey) horizontal lines with $M_{1} < 80$~GeV. 
The most part of the parameter space with singlino-like dark matter
give relatively light LSP with mass in the range $50 - 200$~GeV although
heavier candidates are not entirely excluded. We have found that in
this case, considerable admixture of  higgsinos is always present
(numerically, we obtain $\l(|N_{53}|^{2} +  |N_{54}|^{2}\r)^{(1/2)}  \sim
0.4-0.8$). Singlino dark matter with the mass $m_{\chi}\gsim 80$~GeV
annihilates predominantly into $W^{+}W^{-}$ gauge bosons, while for
$m_{\chi}\sim 0.5M_{h}$ or $0.5M_{Z}$ the main channel is the resonant
one, $\chi\chi\to h^{*}(Z^{0*}) \to f\bar{f}$. One concludes that dark
matter problem can also be solved in the framework of considered
models.  

\section{Discussion and conclusions}
\label{concl_section}
In this paper we proposed a generalization of the minimal split
supersymmetry model by taking into account the necessity to explain
the baryon asymmetry of the Universe. The main idea was to start the
whole construction of split SUSY not with MSSM but with NMSSM. We
needed to fine-tune some parameters of the model to obtain the same
low energy spectrum of particles as in the minimal split SUSY plus
additional singlet particles, which give rise to all features one
needs for successful baryogenesis. We restricted ourselves to rather
small part of parameter space, in which the lightest Higgs boson does
not mix with other scalar particles. Like in the minimal version of
split supersymmetry model, the upper bound on the  mass of the
lightest Higgs boson shifts to larger values in comparison, e.g., with
MSSM.

We have considered the electroweak phase transition and, by exploring
one-loop effective potential at finite temperature, we found that
there is a region in the parameter space where the phase
transition is  strongly first order and the baryon asymmetry is not
washed out after the phase transition is completed. We have used WKB
approximation to calculate the value of the baryon asymmetry. We have
found that this model is indeed capable of producing the right amount
of the baryon asymmetry for realistic values of the width and velocity
of the bubble wall.

We have investigated the contribution of the CP-violating sources into 
the electron and neutron EDMs. As in the minimal split supersymmetry,
their values are close to the present experimental limits and can be
measured in the future experiments hence giving an opportunity
to falsify the suggested solution of the baryon asymmetry problem.

We have explored the dark matter in nonminimal split SUSY. In addition
to the bino and higgsino dark matter candidates there is a region in
the parameter space in which singlino 
can considerably contribute to the LSP state and hence plays the  role
of dark matter. We have found that in the latter case LSP is quite
light 
($m_{\chi} < 150$~GeV) and contains considerable admixture of  
higgsino as well. 

To summarize, the split NMSSM models are capable of solving both
baryon asymmetry and dark matter problems and can be probed by the
next generation of EDM experiments. The collider phenomenology of this
model is quite similar to one of minimal split SUSY, if singlino-neutralino
and higgs-singlet mixing is small. In the opposite case there are
additional 
signatures of this model resembling ones in non-split NMSSM. We leave
the study of LHC prospects in probing this model for the future.

{\bf Acknowledgements.} We thank V.~A.~Rubakov for helpful discussions
and interest in this work. S.~V.~Demidov is indebted to S.~F.~Huber
for discussions. This work was supported in part by the Russian 
Foundation of Basic Research grant 05-02-17363, by grant of the
President of the Russian Federation NS-2184.2003.2, by the grant
NS-7293.2006.2 (government contract 02.445.11.7370) and by fellowships
of the "Dynasty" foundation (awarded by the Scientific Council of
ICFPM). The work of D.G. was also supported by the Russian Foundation
of Basic Research grant 04-02-17448 and by the grant of the President
of the Russian Federation MK-2974.2006.2. Numerical part of the work
was done at the computer cluster in Theoretical Division of INR RAS. 

\section{Appendix A}
\label{apa}
This appendix contains the set of renormalization group equations for
the coupling constants of the model described by the Lagrangian
(\ref{gener_poten}), (\ref{gener_yukava}) at energies below
$m_{s}$. One derives them by making use of the general formulae given
in 
Ref.~\cite{Machacek:1983tz}. 

The 2-loop RGE for gauge couplings have the following form
\begin{gather}
\l(4\pi\r)^{2}\frac{d}{dt}g_{i} = g_{i}^{3}b_{i} + 
\frac{g_{i}^{3}}{\l(4\pi\r)^{2}}
\Bigl(\sum_{j=1}^{3} B_{ij}g_{j}^{2} - \sum_{\alpha=u,d,e}
d_{i}^{\alpha} {\rm Tr}\l(h^{\alpha\dagger}h^{\alpha}\r)\\
- d^{W}_{i}\l(\tilde{g}^{2}_{u} + \tilde{g}^{2}_{d}\r) -
d^{B}_{i}\l(\tilde{g}^{\prime 2}_{u} + 
\tilde{g}^{\prime 2}_{d}\r) - d_{i}^{\lambda}\l(\lambda_{u}^{2}
+ \lambda_{d}^{2}\r) - d_{i}^{\kappa}\kappa^{2}\Bigl),
\nonumber
\end{gather}
where $t = \ln{\bar{\mu}}$ and $\bar{\mu}$ is the renormalization
scale. Here the GUT convention $g_{1}^{2} = (5/3)g^{\prime 2}$ is
used. Below we list the coefficients of $\beta$-functions relevant
within the interval $m_{ew} < \bar{\mu} < m_{s}$. Above the splitting
scale, $m_{s}$, the renormalization group equations become the same as
in the usual NMSSM (see, e.g., Ref.~\cite{King:1995vk}). One has
\be
b = \l(\frac{9}{2}, -\frac{7}{6}, -5\r),\;\;\;
B = 
\l(
\begin{array}{ccc}
\frac{104}{25} & \frac{18}{5} & \frac{44}{5} \\
\frac{6}{5} & \frac{106}{3} & 12 \\
\frac{11}{10} & \frac{9}{2} & 22 \\
\end{array}
\r),
\ee
\be
d^{u} = \l(\frac{17}{10}, \frac{3}{2}, 2\r),\;\;\;
d^{d} = \l(\frac{1}{2}, \frac{3}{2}, 2\r),\;\;\;
d^{e} = \l(\frac{3}{2}, \frac{1}{2}, 0\r),
\ee
\be
d^{W} = \l(\frac{9}{20}, \frac{11}{4}, 0\r),\;\;\;
d^{B} = \l(\frac{3}{20}, \frac{1}{4}, 0\r),\;\;\;
d^{\lambda} = \l(\frac{3}{10}, \frac{1}{2}, 0\r),\;\;\;
d^{\kappa} = \l(\frac{3}{5}, 1, 0\r).
\ee
It is convenient to introduce the following notations
\begin{gather}
T_{H} = {\rm Tr}\left[3h^{u\dagger}h^{u} + 3h^{d\dagger}h^{d} +
h^{e\dagger}h^{e}\right] + \frac{3}{2}\l(\tilde{g}_{u}^{2} +
\tilde{g}_{d}^{2}\r) + 
\frac{1}{2}\l(\tilde{g}_{u}^{\prime 2} + \tilde{g}_{d}^{\prime
  2}\r) + \lambda_{u}^{2} + \lambda_{d}^{2}, \\
T_{N} = 2\l(k^{2} + \kappa^{2}\r).
\end{gather}
One-loop equations for the quark and lepton Yukawa couplings can be
read off from  
Ref. \cite{Giudice:2004tc} with the substitution $T \to T_{H}$. One
obtains 
\be
\l(4\pi\r)^{2}\frac{d}{dt}h^{u} = h^{u}\l(-3
\sum_{i=1}^{3}c_{i}^{u}g_{i}^{2} + \frac{3}{2}h^{u\dagger}h^{u} -
\frac{3}{2}h^{d\dagger}h^{d} + T_{H}\r),
\ee
\be
\l(4\pi\r)^{2}\frac{d}{dt}h^{d} = h^{d}\l(-3
\sum_{i=1}^{3}c_{i}^{d}g_{i}^{2} - \frac{3}{2}h^{u\dagger}h^{u} +
\frac{3}{2}h^{d\dagger}h^{d} + T_{H}\r),
\ee
\be
\l(4\pi\r)^{2}\frac{d}{dt}h^{e} = h^{e}\l(-3
\sum_{i=1}^{3}c_{i}^{e}g_{i}^{2} + \frac{3}{2}h^{e\dagger}h^{e} +
T_{H}\r).
\ee
Here the coefficients $c^{u,d,e}$ are given by
\be
c^{u} = \l(\frac{17}{60}, \frac{3}{4}, \frac{8}{3}\r),\;\;\;
c^{d} = \l(\frac{1}{12}, \frac{3}{4}, \frac{8}{3}\r),\;\;\;
c^{e} = \l(\frac{3}{4}, \frac{3}{4}, 0\r).
\ee
RGE for gaugino 
and singlino couplings (see (\ref{gaugino_couplings}) and
(\ref{singlino_couplings})) are
\begin{gather}
\label{tildegu}
\l(4\pi\r)^{2}\frac{d}{dt}\tilde{g}_{u} = -3\tilde{g}_{u}
\sum_{i=1}^{3}C_{i}g_{i}^{2} + \frac{5}{4}\tilde{g}_{u}^{3} -
\frac{1}{2}\tilde{g}_{u}\tilde{g}_{d}^{2} +
\frac{1}{4}\tilde{g}_{u}\tilde{g}_{u}^{\prime 2} +
\tilde{g}_{d}\tilde{g}_{d}^{\prime}\tilde{g}_{u}^{\prime}
\\
+
\frac{1}{2}\tilde{g}_{u}\tilde{\lambda}_{u}^{2} +
\frac{1}{2}\tilde{g}_{u}\kappa^{2} +
2\tilde{g}_{d}\lambda_{u}\lambda_{d} + \tilde{g}_{u}T_{H},
\nonumber \\
\label{tildeguprime}
\l(4\pi\r)^{2}\frac{d}{dt}\tilde{g}_{u}^{\prime} =
-3\tilde{g}_{u}^{\prime}\sum_{i=1}^{3}C_{i}^{\prime}g_{i}^{2} +
\frac{3}{4}\tilde{g}_{u}^{\prime 3} + 
\frac{3}{2}\tilde{g}_{u}^{\prime}\tilde{g}_{d}^{\prime 2} +
\frac{3}{4}\tilde{g}_{u}^{\prime}\tilde{g}_{u}^{2} +
3\tilde{g}_{u}\tilde{g}_{d}\tilde{g}_{d}^{\prime}
\\
+
\frac{3}{2}\tilde{g}_{u}^{\prime}\tilde{\lambda}_{u}^{2} +
\frac{1}{2}\tilde{g}_{u}^{\prime}\kappa^{2} +
3\tilde{g}_{d}^{\prime}\lambda_{u}\lambda_{d} +
\tilde{g}_{u}^{\prime}T_{H}, 
\nonumber\\
\label{lambdau}
\l(4\pi\r)^{2}\frac{d}{dt}\lambda_{u} =
\frac{3}{2}\lambda_{u}^{3} + 3\lambda_{d}^{2}\lambda_{u} +
2k\lambda_{u} + \frac{1}{2}\lambda_{u}\kappa^{2} +
\frac{1}{4}\lambda_{u}\l(3\tilde{g}_{u}^{2} +
\tilde{g}^{\prime 2}_{u}\r)
\\
+ \lambda_{d}\l(3\tilde{g}_{d}\tilde{g}_{u} +
\tilde{g}_{d}^{\prime}\tilde{g}_{u}^{\prime}\r) - 
3\lambda_{u}\sum_{i=1}^{3}C_{i}^{\prime}g_{i}^{2} +
\lambda_{u}T_{H}, 
\nonumber
\end{gather}
where
\[
C = \l(\frac{3}{20}, \frac{11}{4}, 0\r)\;\;\;
C^{\prime} = \l(\frac{3}{20}, \frac{3}{4}, 0\r)
\]
The equations for $\tilde{g}_{d}$, $\tilde{g}_{d}^{\prime}$ and
$\lambda_{d}$ are the same as (\ref{tildegu}), (\ref{tildeguprime})
and (\ref{lambdau}) with the only replacement $u \leftrightarrow d$. 

RGE for singlino Yukawa couplings $k$ and $\kappa$ are
\be
\l(4\pi\r)^{2}\frac{d}{dt}k = 4k^{3} + 2k\l(\lambda_{u}^{2}
+ \lambda_{d}^{2}\r) - \frac{1}{2}\lambda_{u}\lambda_{d}\kappa + 
kT_{N},
\ee
\be
\l(4\pi\r)^{2}\frac{d}{dt}\kappa = \kappa^{3} +
\frac{1}{2}\kappa\l(\lambda_{2}^{2} + \lambda_{d}^{2}\r) +
\frac{3}{4}\kappa\l(\tilde{g}_{u}^{2} + \tilde{g}_{d}^{2}\r) +
\frac{1}{4}\kappa\l(\tilde{g}_{u}^{\prime 2} +
\tilde{g}_{d}^{\prime 2}\r) 
\ee
\[
+ 4k\lambda_{u}\lambda_{d} - 3\kappa\l(\frac{3}{10}g_{1}^{2} +
\frac{3}{2}g_{2}^{2}\r) + \kappa T_{N}.
\]
RGE for the Higgs quartic coupling is
\be
\l(4\pi\r)^{2}\frac{d}{dt}\tilde{\lambda} = 12\tilde{\lambda}^{2} +
2\kappa_{1}^{2} + 8\kappa_{2}^{2} +
\tilde{\lambda}\Bigl[-9\l(\frac{g_{1}^{2}}{5} + g_{2}^{2}\r) +
  6\l(\tilde{g}_{u}^{2} + \tilde{g}_{d}^{2}\r) +
  2\l(\tilde{g}_{u}^{\prime 2} + \tilde{g}_{d}^{\prime 2}\r)
\ee
\[
+
  4\l(\lambda_{u}^{2} + \lambda_{d}^{2}\r) + 4{\rm
    Tr}\l(3h^{u\dagger}h^{u} + 3h^{d\dagger}h^{d} +
  h^{e\dagger}h^{e}\r)\Bigl] +
\frac{9}{2}\l(\frac{g_{2}^{4}}{2} + \frac{3g_{1}^{4}}{50} +
\frac{g_{1}^{2}g_{2}^{2}}{5}\r)
\]
\[
-4{\rm Tr}\l(3\l(h^{u\dagger}h^{u}\r)^{2} +
3\l(h^{u\dagger}h^{u}\r)^{2} +
\l(h^{u\dagger}h^{u}\r)^{2}\r) - 5\l(\tilde{g}_{u}^{4}
+ \tilde{g}_{d}^{4}\r) - 2\tilde{g}_{u}^{2}g_{d}^{2} 
- \l(\tilde{g}_{u}^{\prime 2} + \tilde{g}_{d}^{\prime 2}\r)^{2} 
\]
\[
- 
2\l(\tilde{g}_{u}\tilde{g}_{u}^{\prime} +
\tilde{g}_{d}\tilde{g}_{d}^{\prime}\r) - 4\l(\lambda_{u}^{2} +
\lambda_{d}^{2}\r)^{2} - 4\l(\lambda_{u}\tilde{g}_{u} +
\lambda_{d}\tilde{g}_{d}\r)^{2} -
4\l(\lambda_{u}\tilde{g}_{u}^{\prime} +
\lambda_{d}\tilde{g}_{d}^{\prime}\r)^{2}.
\]

Finally, RGE for the remaining scalar couplings have the following
form 
\be
\l(4\pi\r)^{2}\frac{d}{dt}\kappa_{1} = 8\lambda_{N}\kappa_{1} +
4\kappa_{1}^{2} + 16\kappa_{2}^{2} + 6\kappa_{1}\tilde{\lambda} -
4\eta\kappa_{2} -8k^{2}\l(\lambda_{u}^{2} + \lambda_{d}^{2}\r)
\ee
\[
+ 
2\kappa_{1}\l(T_{H} + T_{N}\r) -
\frac{9}{2}\kappa_{1}\l(\frac{g_{1}^{2}}{5} + g_{2}^{2}\r),
\]
\be
\l(4\pi\r)^{2}\frac{d}{dt}\kappa_{2} = 6\tilde{\lambda}\kappa_{2} +
8\kappa_{1}\kappa_{2} + 24\xi\kappa_{2} + 4\kappa_{2}\lambda_{N} -
4k\kappa\lambda_{u}\lambda_{d}
\ee
\[
+ 2\kappa_{2}\l(T_{H} + T_{N}\r) -
\frac{9}{2}\kappa_{2}\l(\frac{g_{1}^{2}}{5} + g_{2}^{2}\r), 
\]
\be
\l(4\pi\r)^{2}\frac{d}{dt}\lambda_{N} = 20\lambda_{N}^{2} +
144\kappa_{1}\kappa_{2} +\frac{3}{2}\eta^{2} + 2\kappa_{2}^{2} +
2\kappa_{1}^{2} - 8k^{4} - 2\kappa^{4} + 4\lambda_{N}T_{N}
\ee
\be
\l(4\pi\r)^{2}\frac{d}{dt}\xi = 24\lambda_{N}\xi +
4\kappa_{2}^{2} + \frac{3}{2}\eta^{2} + 4\xi T_{N},
\ee
\be
\l(4\pi\r)^{2}\frac{d}{dt}\eta = 36\eta\lambda_{N} -
12\kappa_{1}\kappa_{2} + 72\xi\eta + 4\eta T_{N}.
\ee

\section{Appendix B}
\label{apb}
This Appendix contains the relevant set of formulas describing the
cross sections of neutralino annihilation into the Higgs bosons. Here
we adopt the notations of Ref.~\cite{Nihei:2002ij} for couplings and
matrices. The relevant part of the Lagrangian is 
\be
{\mathcal L} = \frac{1}{2}\sum_{i,j=1}^{5} \bar{\chi_{i}^{0}}
\l(C_{S}^{\chi_{i}^{0} \chi_{j}^{0} \Phi_{1}}\Phi_{1} + C_{S}^{\chi_{i}^{0}
  \chi_{j}^{0} \Phi_{2}}\Phi_{2}\r) 
\chi_{j}^{0}
 + \frac{1}{2}\l(C^{hh\Phi_{1}}\Phi_{1} + C^{hh\Phi_{2}}\Phi_{2}\r)h^{2},
\ee
where we use $\Phi_{1}, \Phi_{2}$ for mass eigenstates of scalar $S$
and pseudoscalar $P$ fields related to CP-eigenstates as follows
\[
\l(
\begin{array}{c}
S \\
P
\end{array}
\r)
=
\l(
\begin{array}{c}
\cos{\theta}\Phi_{1} + \sin{\theta}\Phi_{2} \\
-\sin{\theta}\Phi_{1} + \cos{\theta}\Phi_{2}
\end{array}
\r)
\]
with a mixing angle $\theta$ and
\be
C_{S}^{\chi_{i}^{0}\chi_{j}^{0}\Phi_{1}} = -\sqrt{2}k N^{*}_{i5}
N^{*}_{j5}{\rm e}^{-i\theta},\;\;\; 
C_{S}^{\chi_{i}^{0}\chi_{j}^{0}\Phi_{2}} = -i\sqrt{2}k N^{*}_{i5}
N^{*}_{j5}{\rm e}^{-i\theta},
\ee
\be
C^{hh\Phi_{1}} = \frac{1}{\sqrt{2}}\l(\tilde{A}_{2}\cos{\theta} +
\tilde{A}_{1}\sin{\theta}\r) + \frac{1}{2}\l[\l(\kappa_{1} -
  \kappa_{2}\r)v_{S}\cos{\theta} - \l(\kappa_{1} + \kappa_{2}\r)v_{P}\sin{\theta}\r],
\ee
\be
C^{hh\Phi_{2}} = \frac{1}{\sqrt{2}}\l(\tilde{A}_{2}\sin{\theta} -
\tilde{A}_{1}\cos{\theta}\r) + \frac{1}{2}\l[\l(\kappa_{1} -
  \kappa_{2}\r)v_{S}\sin{\theta} + \l(\kappa_{1} + \kappa_{2}\r)v_{P}\cos{\theta}\r]
\ee
For simplicity we assume, as everywhere in this paper, that the Higgs
state $h$ does not mix with other (pseudo)scalar particles. We present
functions $\tilde{w}_{hh}$, determining the cross sections of
nonrelativistic scattering $\chi\chi\to hh$ (see
Ref.~\cite{Nihei:2002ij} for definitions of these and other related
quantities). There are contributions to the cross sections coming from
the s-channel Higgs ($h$), scalar ($S$) and pseudoscalar ($P$)
exchanges and $t$- and $u$-channel neutralino exchange as well as from
their interference,
\be
\tilde{w}_{hh} = \tilde{w}_{hh}^{(h,S,P)} +
\tilde{w}_{hh}^{(\chi^{0})} + \tilde{w}_{hh}^{(h,S,P-\chi^{0})}.
\ee
For scalar exchanges in the $s$-channel one has
\be
\tilde{w}_{hh}^{(h,\Phi_{1},\Phi_{2})} = \frac{1}{2}
\left|\sum_{r = h,\Phi_{1},\Phi_{2}}\frac{C^{hhr}C_{S}^{\chi\chi r}}{s - m_{r}^{2} +
  i\Gamma_{r}m_{r}}\right|^{2}\left(s - 4m_{\chi}^{2}\right),
\ee
The scalar - neutralino interference gives
\be
\tilde{w}_{hh}^{(h,\Phi_{1},\Phi_{2}-\chi^{0})} = 
2\sum_{i=1}^{5}{\rm Re} 
\left[\sum_{r=h,\Phi_{1},\Phi_{2}}\l(\frac{C^{hhr}C_{S}^{\chi\chi
      r}}{s - m_{r}^{2} +
    i\Gamma_{r}m_{r}}\r)^{*}C^{\chi_{i}^{0}\chi 
    h}_{S}C^{\chi_{i}^{0}\chi h}_{S}\right]
\ee
\[
\times \left\{-2m_{\chi} + [2m_{\chi}(m_{h}^{2}-m_{\chi_{i}^{0}}^{2} -
    m_{\chi}^{2})
    + m_{\chi_{i}^{0}}(s - 4m_{\chi}^{2})]
{\mathcal F}(s,m_{\chi}^{2},m_{h}^{2},m_{h}^{2},m^{2}_{\chi_{i}^{0}})\right\}
\]
The term $\tilde{w}_{hh}^{\chi^{0}}$ corresponding to $t$- and
$u$-channel neutralino exchange is 
the same as in Ref.~\cite{Nihei:2002ij}.

\bibliographystyle{asmplain}



\end{document}